\documentclass[preprint,secnumarabic,amssymb,superscriptaddress,nobibnotes,aps,showpacs,nofootinbib,noeprint]{revtex4-1}
\usepackage[english]{babel}
\usepackage{graphicx}
\usepackage{pgf}

\usepackage{textcomp}

\usepackage{amsmath}
\usepackage{amsfonts}
\usepackage{amssymb}
\usepackage{mathtools}
\usepackage{siunitx}

\usepackage{hyperref}
\usepackage{rotating}
\usepackage{booktabs} 
\usepackage[sort&compress]{cleveref}
\usepackage{tocbasic}
\usepackage{slashed}
\usepackage{transparent}

\DeclareMathAlphabet{\mathbb}{U}{bbold}{m}{n}

\usepackage{ifthen}

\newcommand{\hideandshow}[1]{
 \ifthenelse{\isundefined{\showme}}{}{#1}}

\begin{document}

\title{Hard sphere crystal nucleation rates: Reconciliation of simulation and experiment}

\author{Wilkin W{\"oh}ler}
\email{Wilkin.Woehler@ise.fraunhofer.de}
\affiliation{Institute of Physics, University of Freiburg, Hermann-Herder-Str. 3, 79104 Freiburg, Germany}
\author{Tanja Schilling}
\email{Tanja.Schilling@physik.uni-freiburg.de}
\affiliation{Institute of Physics, University of Freiburg, Hermann-Herder-Str. 3, 79104 Freiburg, Germany}

\date{\today}

\begin{abstract}
  Over the past two decades, a large number of studies addressed the topic of crystal nucleation in suspensions of hard spheres. The shared result of all these efforts is that, at low super-saturations,  experimentally observed nucleation rates and numerically computed ones differ by more than ten orders of magnitude.
We present precise simulation results of crystal nucleation rate densities in the meta-stable hard sphere liquid. To compare these rate densities to experimentally measured ones, we propose an interpretation of the experimental data as a combination of nucleation and crystal growth processes (rather than purely the nucleation process). This interpretation may resolve the long standing dispute about the differing rates.
\begin{center}
\rule{0.9\textwidth}{0.5pt}
\end{center}
\end{abstract}

\maketitle

Precise predictions of crystallization and solidification kinetics are highly sought after in scientific fields like atmospheric physics, metallurgy and chemical engineering. Arguably, the simplest model to study the liquid-to-crystal transition is the hard sphere fluid. 
Following the seminal computer simulation study by Alder and Wainwright in 1955 \cite{Alders59} and the resourceful experiments by Pusey and van Megen in 1987 \cite{Pusey1986}, the crystal nucleation process in suspensions of colloidal hard spheres has been the topic of extensive research in the past decades. Strikingly, despite the simplicity of the system, there are pronounced differences between theoretically predicted and experimentally measured nucleation rate densities, as remarked by Auer and Frenkel in 2001 \cite{Auer2001}. This observation has been confirmed repeatedly, and many authors emphasized their astonishment about the unexplained discrepancy of 10 orders of magnitude between theory and experiment \cite{Filion2010,Filion2011,russo2013,russo2016,wood2018}.

As causes for the discrepancy on the experimental side have been suggested the size polydispersity of the colloidal particles \cite{Auer2001}, their electrostatic charge \cite{auer2002}, hydrodynamic coupling between the colloids due to the solvent \cite{Radu2014,Tateno2019,Fiorucci2020}, sedimentation in the samples \cite{russo2013,wood2018}, and the fact that it is difficult to control the volume fraction \cite{Pusey2009,Poon2012,Royall2013}. However, although these effects are relevant to a certain degree, neither resolved the disagreement. On the theoretical side, the choice of sampling technique has been suggested \cite{Filion2010}, the choice of reaction coordinate \cite{Sear2012,Schilling2011} and the validity of the assumption, that the nucleation process were Markovian \cite{Kuhnhold2019,Meyer2021}. Again, neither of these could explain a discrepancy of such a staggering magnitude.

\begin{figure}[h]
\centering
\includegraphics[width=0.95 \linewidth]{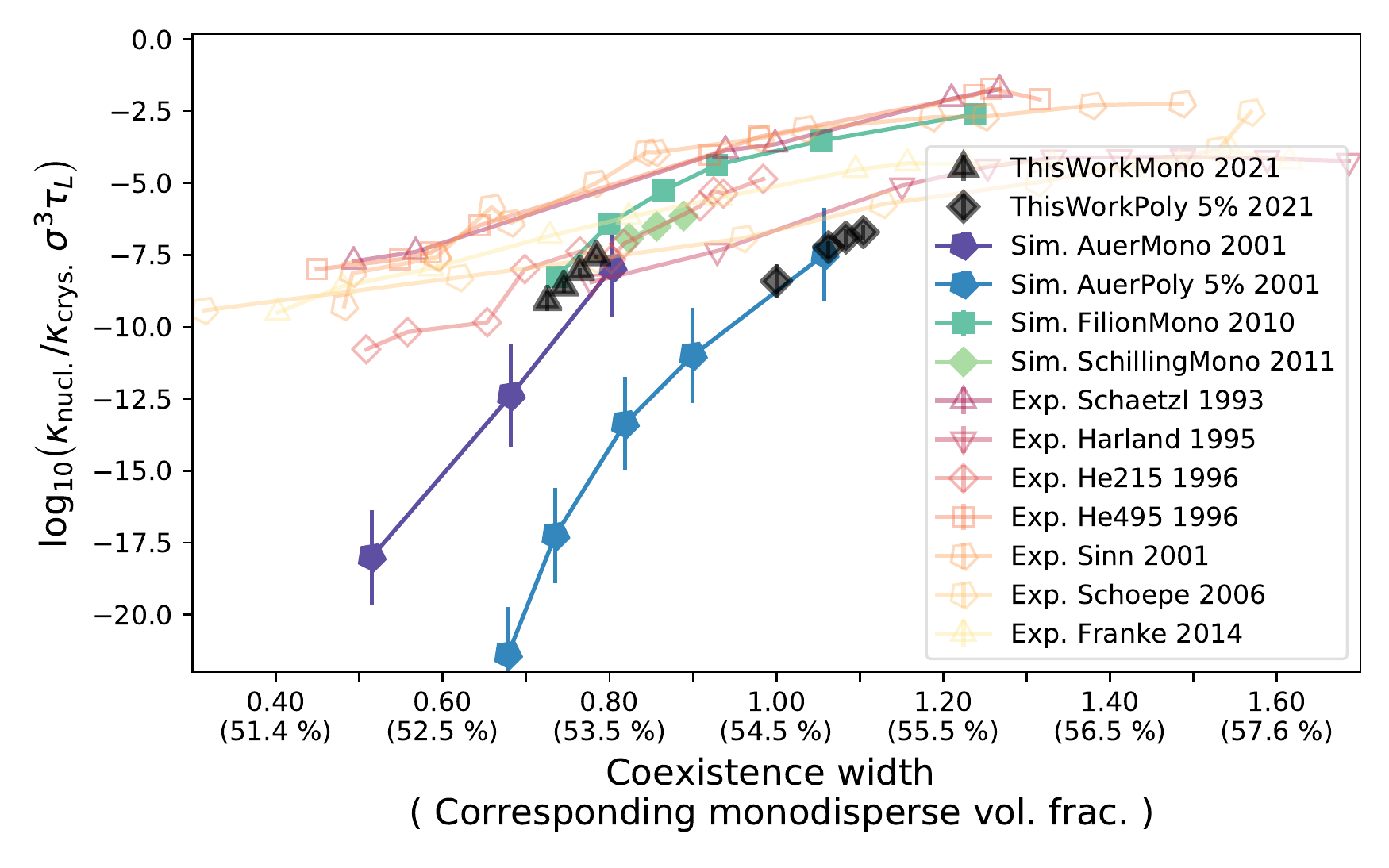}
\caption[]{Crystal nucleation rates at different volume fractions. Experimental data (open symbols, orange and red) and numerical data (closed symbols, green and blue) from refs.~\cite{Harland1997,He1996,schaetzel1993,Sinn2001,Auer2001,Filion2010a,Fiorucci2020a,Schilling2011}, as well as this work (black symbols, see also \autoref{tab:nuc_rate_mono} and \autoref{tab:nuc_rate_poly}). 
}
\label{fig:nucleation_comparison}
\end{figure}

The central ``bone of contention'' is shown in fig.~\ref{fig:nucleation_comparison}, which presents the crystal nucleation rate density as a function of the volume fraction of spheres. As unit of length we used the particle diameter $\sigma$ and as unit of time the long-time self-diffusion time $\tau_L = (6 D_L^S)^{-1}$, where $D_L^S$ is the long-time self-diffusion coefficient. This choice of units is common in the literature, because it allows for a consistent comparison of the rather varied experimental realizations of the system. The striking point about \autoref{fig:nucleation_comparison} is the difference in slope between the experimental data and the numerical data. While the effects listed above could explain uncertainties in the location of some of the curves, none explained the significant difference in slope.  

To solve this issue, we investigated nucleation and crystal growth by means of an efficient event driven molecular dynamics simulation code (EDMD), which we developed based on the algorithm proposed by Bannerman in 2014 \cite{Bannerman2014}. In this letter, we first present a precise computation of the homogeneous nucleation rate density. Then we recall the methods by which hard sphere nucleation experiments are evaluated and use the Johnson-Mehl-Avrami-Kolmogorow model (JMAK model) to compare our data to the experiments. In particular, we take into account the fact that polycrystalline clusters form by a successive twinning process. We conclude that the experimental data presented in fig.~\ref{fig:nucleation_comparison} is not the crystal nucleation rate but the rate at which mono-crystalline domains form. Thus we argue that the discrepancy between experiment and theory is due to a misinterpretation of the experimental data.

The simulated system consists of up to a few million mono- or polydisperse hard particles enclosed by cubic periodic boundary conditions. The particles' (average) diameter $\sigma$, mass $\text{m}$, and the thermal energy $\text{k}_\text{B} \text{T}$ are chosen as units resulting in a unit of time of $\delta t = \sqrt{m / (k_B T)} \, \sigma $. 
For nucleation measurements, the system is rapidly compressed from the stable liquid state to a meta-stable volume fraction.

We monitor the crystallinity of the system in terms of the q6q6-bond-order parameters \cite{Steinhardt1983, TenWolde1995}, details are given in \cite{Meyer2021}. As the main observable we measure the number of particles within the largest crystalline cluster.
The time between the volume quench and the first nucleation event (the ``nucleation time'') is determined by following the trajectory of the largest cluster within a crystallized system back to the point in time when its size last crossed the size of the average largest cluster of the meta-stable fluid.
As the nucleation time increases drastically for low supersaturation, it is not feasible to run a set of simulations in which the complete ensemble ends in the crystalline state. Therefore, a definition of the nucleation rate, that does not depend on a fully crystallized simulation ensemble, is required.

Here, we define the nucleation rate via the assumption of exponentially distributed nucleation times. 
For each measured distribution, we use the maximum likelihood estimator (M.L.) of a censored exponential distribution \cite{Deemer1955}.

\begin{figure}[!h]
\centering
\includegraphics[width=0.95\linewidth]{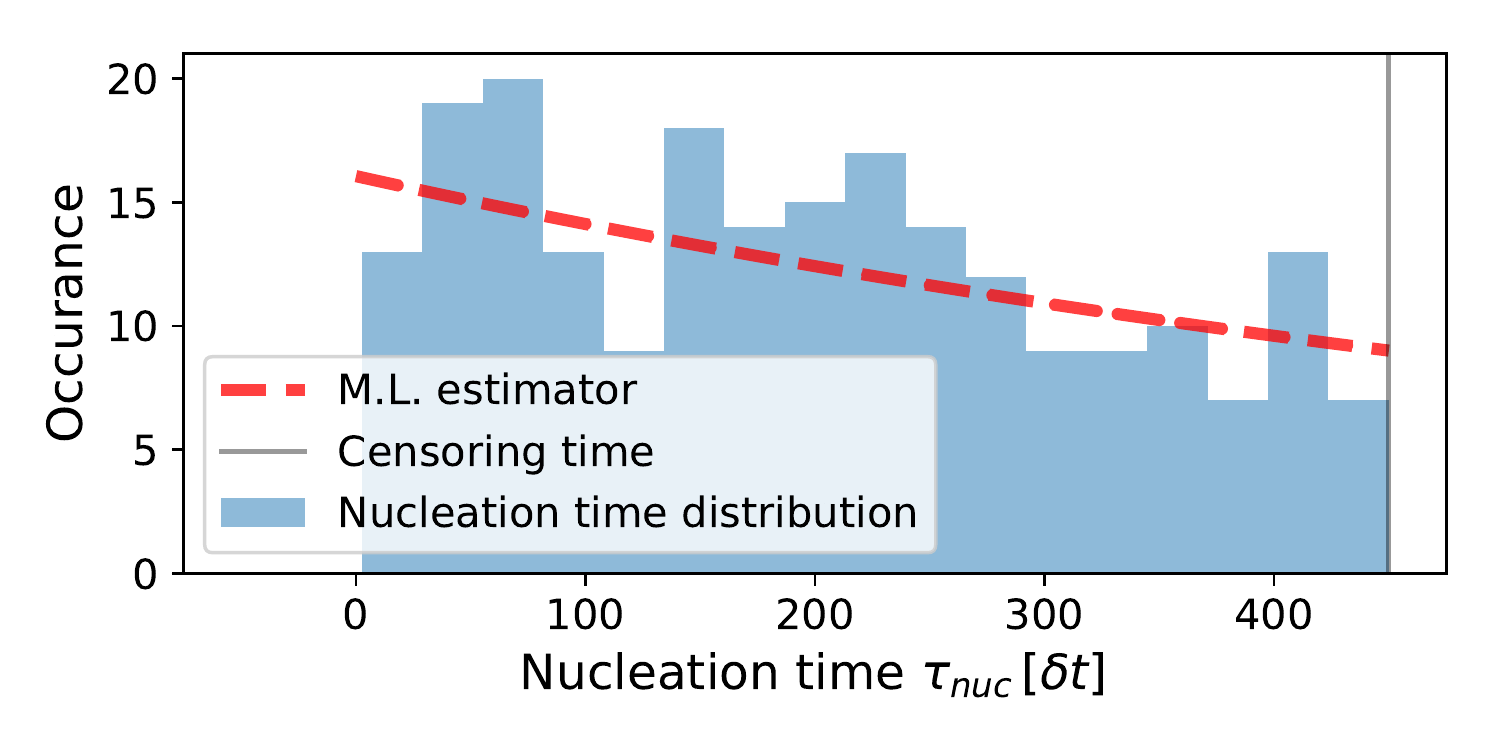}
\caption[]{Nucleation time distribution for 219 crystallization events out of 500 trajectories.}
\label{fig:nucleation_distributions}
\end{figure}

An example of a nucleation time distribution with the corresponding maximum likelihood estimation is shown in \autoref{fig:nucleation_distributions}. The example is measured in a mono-disperse system with a volume fraction of $\eta=\SI{0.534}{}$. We observe, that the simple exponential model describes the distribution well, indicating that the assumption of a time-independent nucleation rate is justified.

To determine the nucleation rates, we ran 500 trajectories of mono-disperse samples at each volume fraction \mbox{$\eta \in \{ \SI{0.531}{}, \SI{0.532}{}, \SI{0.533}{}, \SI{0.534}{}\} $}, where each simulation consisted of $\SI{1048576}{\text{particles}}$. Furthermore, at each volume fraction \mbox{$\eta \in \{ \SI{0.555}{},\SI{0.554}{},\SI{0.553}{},\SI{0.550}{} \} $} we carried out 100  simulations, each consisting of $\SI{16384}{\text{particles}}$ with 5\% Gaussian poly-dispersity. The analysis of the poly-disperse samples relies on the absence of competing phase transitions, but, as remarked by Bolhuis and Kofke \cite{Bolhuis1996} fractionation effects only occur above volume fractions of \SI{5}{\%}. As we have not seen any delay in the nucleation times after the quench, our simulations confirm that fractionation is not significant. 
For poly-disperse systems, it does not make sense to plot the rates versus the plain volume fraction, as the coexistence region is shifted with respect to the one of mono-disperse hard spheres. In principle, it would be desirable to plot all data as a function of the difference in chemical potential between the meta-stable liquid and the stable crystal. However, as it is time-consuming to compute this difference, we use instead
\begin{align}
 u(\eta) = \frac{\eta - \eta_{\text{freeze}}}{\eta_{\text{melt}} - \eta_{\text{freeze}}} \; \text{,}
\end{align}
which is close to the mapping given by the chemical potential, much simpler and frequently used in experimental analyses. 
The poly-disperse transition points $\eta_{\text{freeze}}$ and $\eta_{\text{melt}}$  are taken from ref.~\cite{Bolhuis1996}.

Our results, Tab.~\ref{tab:nuc_rate_mono} and Tab.~\ref{tab:nuc_rate_poly}, agree well with older, less accurate simulation results. We conclude that the mismatch between experiments and simulations is not due to the choice of simulation method. Thus we revisit the analysis of the experimental data.

\begin{table}
\begin{tabular}{c|c}
Volume fraction $ \eta $& Nucleation rate density $\kappa$  [$\sigma^{-3}\tau_{L}^{-1}$]\\ \hline
0.531 & $7.69 \cdot 10^{-10} \pm 2.55 \cdot 10^{-10} $\\ \hline
0.532 & $2.46 \cdot 10^{-9} \pm 4.62 \cdot 10^{-10} $\\ \hline
0.533 & $8.59 \cdot 10^{-9} \pm 9.22 \cdot 10^{-10} $\\ \hline
0.534 & $2.78 \cdot 10^{-8} \pm 1.90 \cdot 10^{-9} $\\ \hline
\end{tabular}
\caption[]{Nucleation rate densities of the mono-disperse hard sphere fluid. 
}
\label{tab:nuc_rate_mono}
\end{table}

\begin{table}
\begin{tabular}{c|c}
Volume fraction $ \eta $& Nucleation rate density $\kappa$  [$\sigma^{-3}\tau_{L}^{-1}$]\\ \hline
0.550 & $2.44 \cdot  10^{-9} \pm 1.24 \cdot 10^{-9} $\\ \hline
0.553 & $5.94 \cdot  10^{-8} \pm 2.17 \cdot  10^{-8} $\\ \hline
0.554 & $1.29 \cdot  10^{-7} \pm 3.99 \cdot  10^{-8} $\\ \hline
0.555 & $1.97 \cdot  10^{-7} \pm 5.05 \cdot  10^{-8} $\\ \hline
\end{tabular}
\caption[]{Nucleation rate densities of a \SI{5}{\%} Gaussian poly-disperse hard sphere fluid.}
\label{tab:nuc_rate_poly}
\end{table}

\begin{figure}[h]
\centering
\includegraphics[width=0.8 \linewidth]{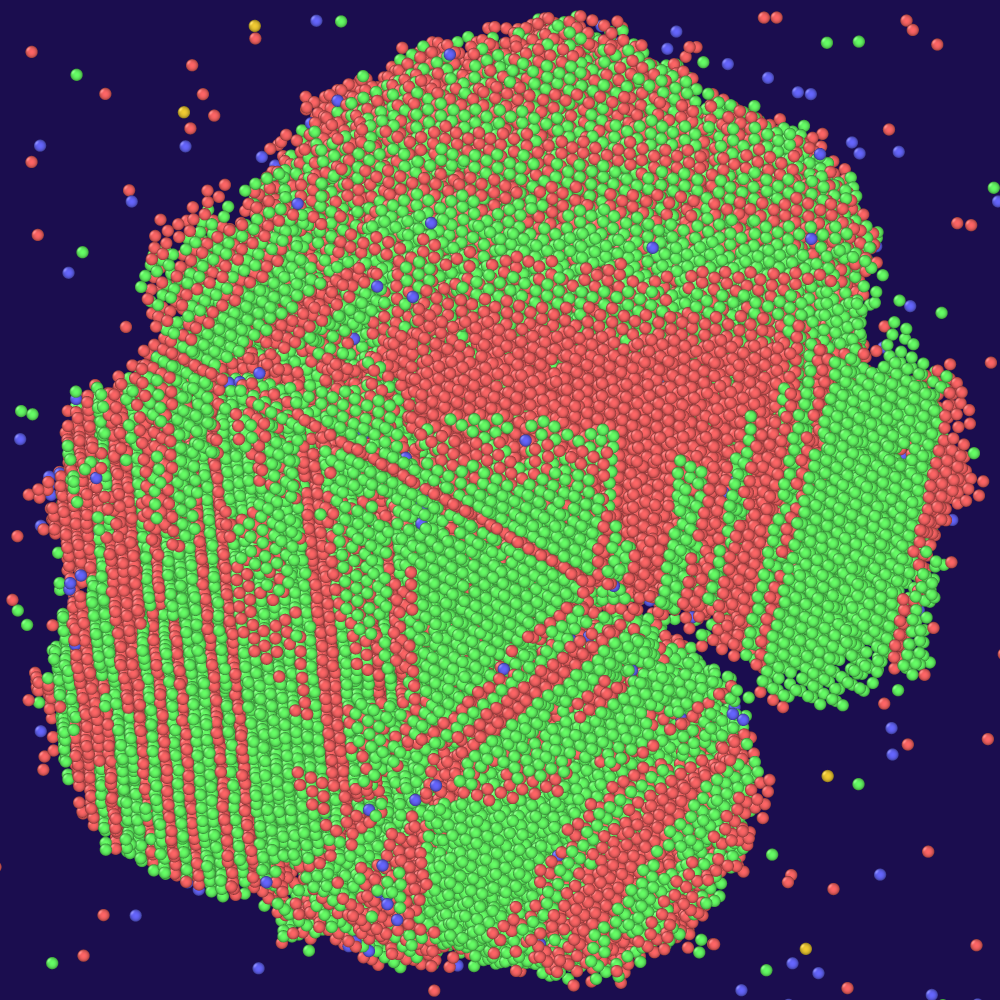}
\caption[]{Cross section through a snapshot of a crystallite. Particles with face centered cubic environments are marked in green, hexagonally closed packed in red, blue is bcc, and particles marked in orange have an icosahedral surrounding. Particles without a crystalline environment are not shown. The visualization is done by the OVITO-program\cite{ovito}}
\label{fig:snapshot}
\end{figure}

In fig.~\ref{fig:snapshot} we show a slice through a crystallite obtained in our simulations. The colours encode crystalline structures, particles in the liquid are not shown. Clearly the cluster is poly-crystalline and twinned. Other snapshots (see SI) show icosahedral domains in addition. We observed only very few mono-crystals. Poly-crystalline clusters produced by a successive twinning process have also been observed by O'Malley and Snook \cite{omalley2003}. We will argue in the following that this observation contradicts one of the basic assumptions made in the analysis of the experimental data, and that this is the cause of the discrepancy in nucleation rates. 

Hard sphere nucleation rates have been determined in three different types of experiments: Bragg scattering (reported e.g.~in ref.~\cite{harland1995,Sinn2001,schope2006,iacopini2009}), small angle light scattering (reported e.g.~in ref.~\cite{schaetzel1993,schatzel1993a,He1996,Sinn2001}), and confocal microscopy (reported by Gasser et al.~\cite{gasser2001}). 

The system analysed by Gasser contained $4000$ particles in the microscope's field of view, of which ca.~200 formed crystallites. As these crystallites were too small to form crystalline domains, we focus our discussion on the scattering experiments. 

In ref.~\cite{harland1995,Sinn2001,schope2006,iacopini2009} the Bragg scattering data was analysed as follows: 
The structure factor of a crystallizing sample was measured as a function of time. Then the structure factor of the liquid phase was subtracted in order to obtain only the signal that stemmed from the crystalline components. The degree of crystallinity $\chi(t)$ was then determined based on the area under the main Bragg peak of the crystalline structure factor, and the average linear size $L(t)$ of the crystallites was determined based on the width of the peak at half maximum (according to the Scherrer equation). Finally, the number of crystallites was computed as $N_c(t) = \chi(t)/L(t)^3$. 

To our knowledge, this type analysis was first presented for hard spheres by Harland and van Megen in the 1990s, and all analyses carried out in more recent work follow their protocol. Harland and van Megen assumed that $L(t)$ was the size of individual crystals rather than domains, although domains produce a Bragg peak of the same shape. We do not know why they excluded the possibility of poly-crystalline growth.  

\begin{figure}
    \includegraphics[width=0.9 \linewidth]{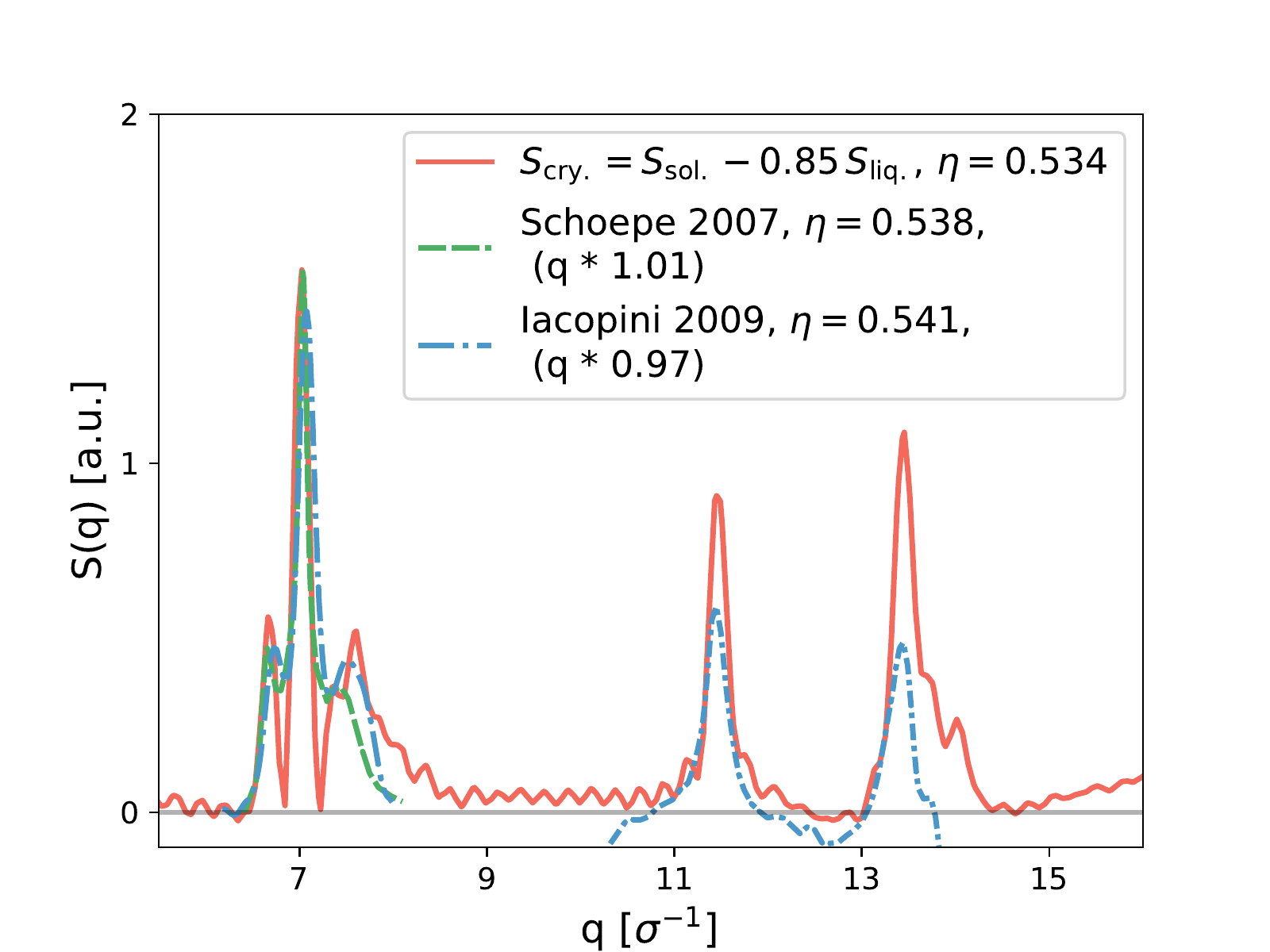}
    \caption[]{Crystalline part of structure factor $S$ as a function of wave number $q$, obtained from simulation snapshots and published experimental data. See main text for details.}
    \label{fig:sf_sim_exp_comparison}
\end{figure}

To check whether our simulation data is consistent with the published Bragg scattering signals, we have produced "scattering data" from simulated configurations, i.e.~we computed the radial distributions functions of the crystallizing systems and Fourier transformed them. 
Then we subtracted the liquid background. The resulting structure factor is shown in fig.~\ref{fig:sf_sim_exp_comparison} (solid line). For comparison we show the experimental structure factors from Sch\"ope \cite{schope2006} (dashed line) and Iacopini \cite{iacopini2009} (dash-dotted line). 
As we do not know the crystallinity exactly, we scaled the structure factors to match the heights of the first Bragg peaks, thus only the relative intensities between the peaks are meaningful. Furthermore, we multiplied the $q$-data of ref.~\cite{schope2006} by a factor of 1.01 and the one of ref.~\cite{iacopini2009} by 0.97, which is compatible with the uncertainty of measured sphere radii. 
The resulting signals are nearly identical. Hence we conclude that our simulations, which show poly-crystalline growth, are consistent with the Bragg scattering signals from experiments in the literature.\\

Regarding the small angle scattering experiments (SAS), we cannot show a direct comparison, because our simulations do not allow us to access sufficiently large length scales. Thus we base our arguments on ref.~\cite{schaetzel1993}: The SAS signal stems from spatial variations in the refractive index, which are caused by spatial variations in the colloid concentration. Ackerson and Sch\"atzel assumed that the signal was produced by the drop in concentration at the crystal-liquid interface \cite{schaetzel1993}. However, they observed that the signal continued to evolve in time after they expected crystal growth  to have stopped. They attributed this phenomenon to coarsening, i.e.~to the dynamics of the grain boundaries, at which there is also a drop in concentration. According to Babinet's theorem the crystal-liquid interface and the amorphous slit between two crystalline domains produce the same scattering pattern. Hence it makes sense that crystallites embedded in a liquid and grain boundaries in the crystal produce similar SAS signals. We conclude that the SAS signals are also consistent with the assumption of poly-crystalline growth.\\

To interpret the nucleation measurements, we use the JMAK-model \cite{Kolmogorow1937,Avrami1939,Avrami1940,Avrami1941}, which describes a crystallization process with stochastically distributed, independent nucleation events and spherically growing clusters. Here we consider the case of a time-independent homogeneous nucleation rate density, $\kappa_{\text{hom}}$, and a time-independent radial growth rate of the clusters, $c$ (for a justification of the latter assumption see SI).
According to the JMAK-model, the fraction of the system's volume which is filled by crystals increases as
\begin{align}
\label{eqn:Avrami_crystallinity}
X_s(t) = 1 - \exp \left( - \frac{\pi}{3} c^3 \kappa_{\text{hom}} t^4 \right) \; \text{.}
\end{align}
(See SI for a brief derivation of this equation.)
Thus the {\it characteristic filling time} $t^*_{p}$ at which a fraction $p$ of the whole system has undergone crystallization is given by
\begin{align}
\label{eqn:filling_time_1}
  t^*_{p} &= \sqrt[4]{- \frac{3 \text{log}(1-p)}{\pi c^3 \kappa_{\text{hom}}}} \; \text{.}
\end{align}
For the following analysis we need to define a time that can be compared to the experimental induction time. We define a characteristic time $t^*$ at which crystallization is in full progress
\begin{align}
\label{eqn:filling_time_2}
 t^* &\coloneqq \left.t^*_{p} \right|_{p = 1 - \exp \left( -\frac{\pi}{3}\right)} =\sqrt[4]{ \frac{1}{c^3 \kappa_{\text hom}}} \; \text{.}
\end{align}
As experiments, simulations and the JMAK-model alike show a steep increase of the crystallinity after an initially slow growth phase, the specific choice of $p$ does not influence $t^*$ by much. Therefore, the results obtained by setting it to the computationally convenient value $p= 1 - \exp \left( -\frac{\pi}{3}\right)$ do not differ much from those obtained with other values of $p$.

Prior to events in which clusters merge, the total number of mono-crystalline domains is given by the sum of the homogeneously nucleated clusters and of the mis-oriented domains grown on the cluster surface. In \autoref{eqn:rate_sum} we write this relation in terms of rates, with the total mono-crystalline domain rate density, $\kappa_{\text{mono}}$, and the rate density for the growth of mis-oriented domains on the interface, $\kappa_{\text{domain}}$. (We average over the rates for events on differently orientated crystalline planes and for different twinning processes.)
\begin{align}
\label{eqn:rate_sum}
\kappa_{\text{mono}}(t) = \kappa_{\text{hom}} + \frac{ A_{\text{crys}}(t) \kappa_{\text{domain}}}{V} \quad ,
\end{align}
where $A_{\text{crys}}$ is the interfacial area between the crystallites and the liquid and $V$ is the system volume, i.e.
\begin{equation}
\frac{A_{\text{crys}}}{V} (t) \approx 
                                                                                                                  \int_0^t  \kappa_{\text{hom}}\, 4 \pi c^2 (t-\tau)^2 d \tau 
= \frac{4 \pi}{3} \kappa_{\text{hom}} c^2 t^3 \; \text{.}\label{eqn:solid_area3}
\end{equation}
Inserting \autoref{eqn:solid_area3} into \autoref{eqn:rate_sum} we obtain 
\begin{align}
\label{eqn:k_crys_2}
  \kappa_{\text{mono}}(t) = \kappa_{\text{hom}} \left(1 + \frac{4 \pi}{3} \kappa_{\text{domain}} c^2 t^3 \right) \; \text{.}
\end{align}
To compare $\kappa_{\text{mono}}$ to the experimental data, we evaluate \autoref{eqn:k_crys_2} at  the characteristic time $t^*$ from \autoref{eqn:filling_time_2}.

\begin{align}
\label{eqn:domain_nuc_rate_density}
\left. \kappa_{\text{mono}} \right|_{t = t^*} &= \kappa_{\text{hom}} + \frac{4 \pi}{3} \sqrt[4]{\frac{\kappa_{\text{hom}}}{c}} \kappa_{\text{domain}}
\end{align}
We determined the value of $\kappa_{\text{domain}}$ by visual inspection of our simulation snapshots to be $\kappa_{\rm domain} = 5 \cdot 10^{-4} (\sigma^2 \, \tau_{\rm L})^{-1}$ (see SI for a detailed description of the procedure).

\begin{figure}[]
\centering
\includegraphics[width=0.95 \linewidth]{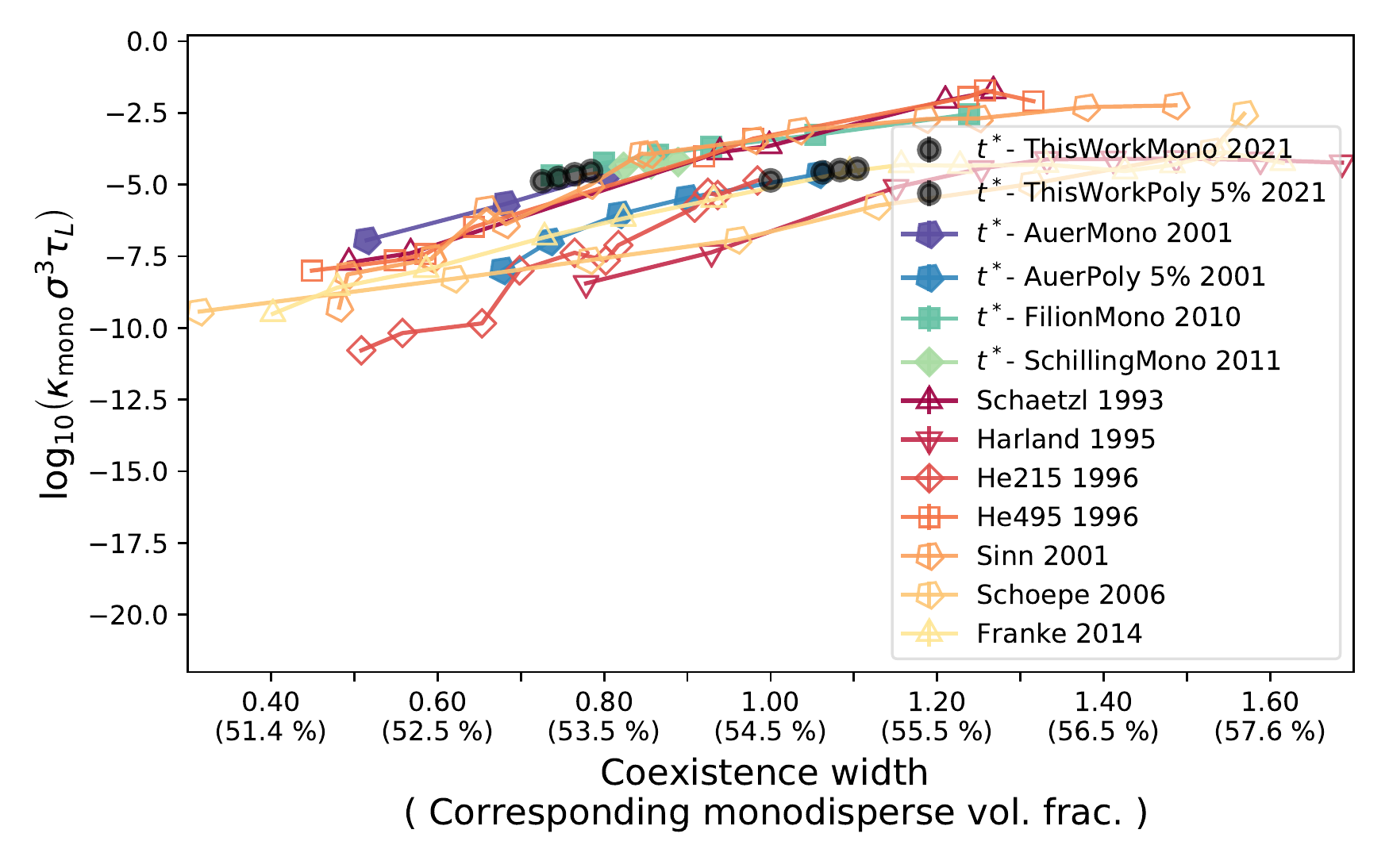}
\caption[]{Rate of formation of mono-crystalline domains \autoref{eqn:domain_nuc_rate_density} at a characteristic time $t^*$. The gap between experiment and simulation is closed.}
\label{fig:mod_nucleation_comparison_3}
\end{figure}

 We suppose that the rate that has been determined in Bragg-scattering experiments is $\kappa_{\text{mono}}$ rather than the bare homogeneous nucleation rate $\kappa_{\text{hom}}$. In \autoref{fig:mod_nucleation_comparison_3} we show the same experimental data as \autoref{fig:nucleation_comparison}, but all simulation data is now given in terms of $\kappa_{\text{mono}}$. The ten orders of magnitude gap is closed and there is remarkable agreement between the simulations and the experiments.

The agreement is particularly clear, if the poly-dispersity of the experimental samples is taken into account. The larger the poly-dispersity, the lower the rate, i.e.~the simulation data on the top of the bundle of lines, which stems from mono-disperse systems and the experimental data on the bottom, which stems from  samples with more than \SI{5}{\%} poly-dispersity, are consistent with our interpretation.

In summary, we analysed and characterized the nucleation and crystallization process of the meta-stable hard sphere system in detail. We calculated nucleation rate densities from simulation data by means of a  Maximum Likelihood estimator and a Monte Carlo uncertainty propagation. The results of our EDMD simulations agree with previous results, but have a quantified and higher precision.

To compare the simulation data to experiments, we invoke the JMAK-model of crystal nucleation and growth. We propose to interpret the results from scattering experiments as a combination of homogeneous nucleation rates and the rate at whichmis-oriented domains grow in a successive twinning process at the cluster surface.
Our central result is the reconciliation of simulation studies and experimental measurements concerning the nucleation rate in the meta-stable hard sphere liquid. We conclude, that earlier comparisons failed, because different quantities were compared under the same name.  The mismatch of names seems to result mostly from interpreting the shape of scattering signals as originating from full clusters rather than mono-crystalline domains.

The authors acknowledge support by the state of Baden-Wuerttemberg through bwHPC
and the German Research Foundation (DFG) through grant no.~INST 39/963-1 FUGG (bwForCluster NEMO) and through grant no.~430195928. H.~J.~Sch\"ope kindly provided the experimental data sets as a function of the normalized coexistence width. We thank H.~J.~Sch\"ope, T.Palberg and S.~Egelhaaf for feedback on the manuscript.

\appendix
\section{Snapshots of Crystallites}

\begin{figure}[h]
\begin{center}
    \includegraphics[width=0.3\linewidth]{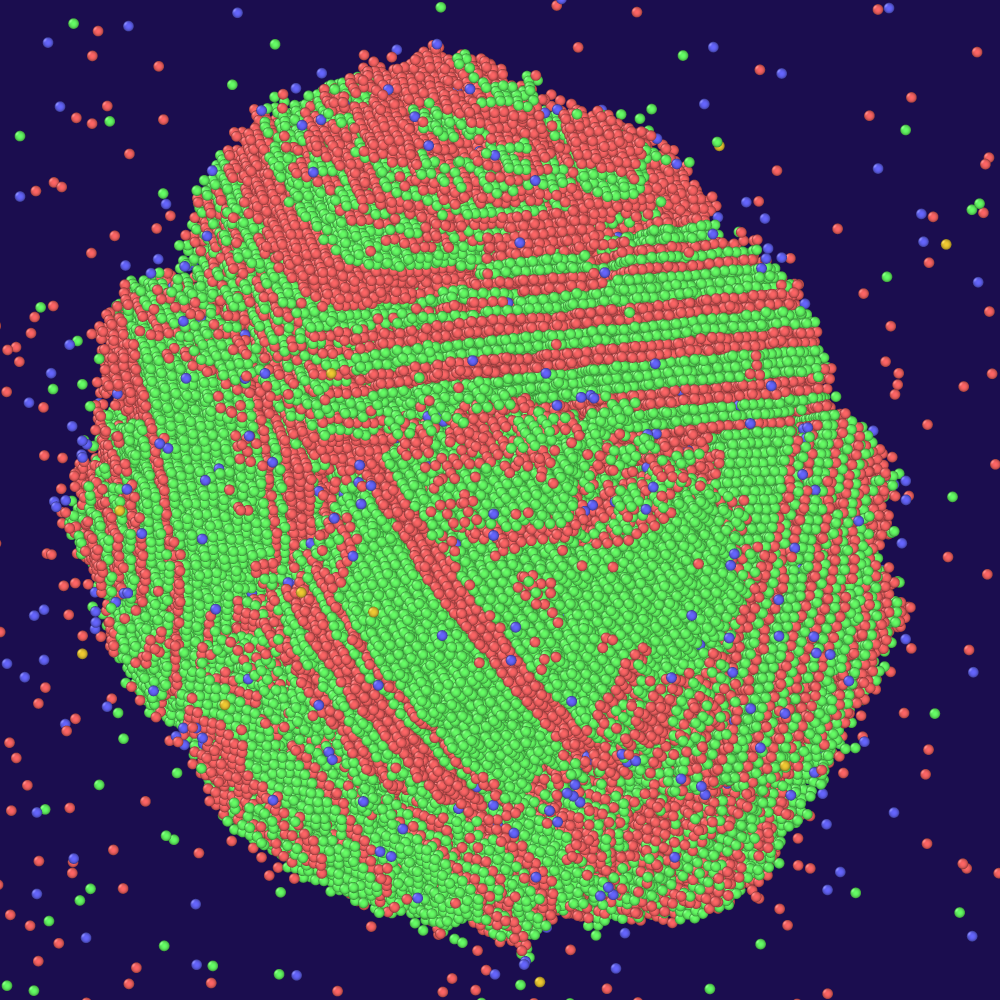}
    \includegraphics[width=0.3\linewidth]{fcc_332_slice.png}
\end{center}
\caption{Example of an fcc/hcp cluster with twinning, $\eta=0.534$, Left: Cluster seen from the outside, right: Slice through the cluster. Red corresponds to hcp and green to fcc stacking, blue is bcc, and particles marked in orange have an icosahedral surrounding, particles in a liquid neighbourhood are not shown.}
\label{fig:snapshot1}
\end{figure}

\begin{figure}[h]
\begin{center}
    \includegraphics[width=0.3\linewidth]{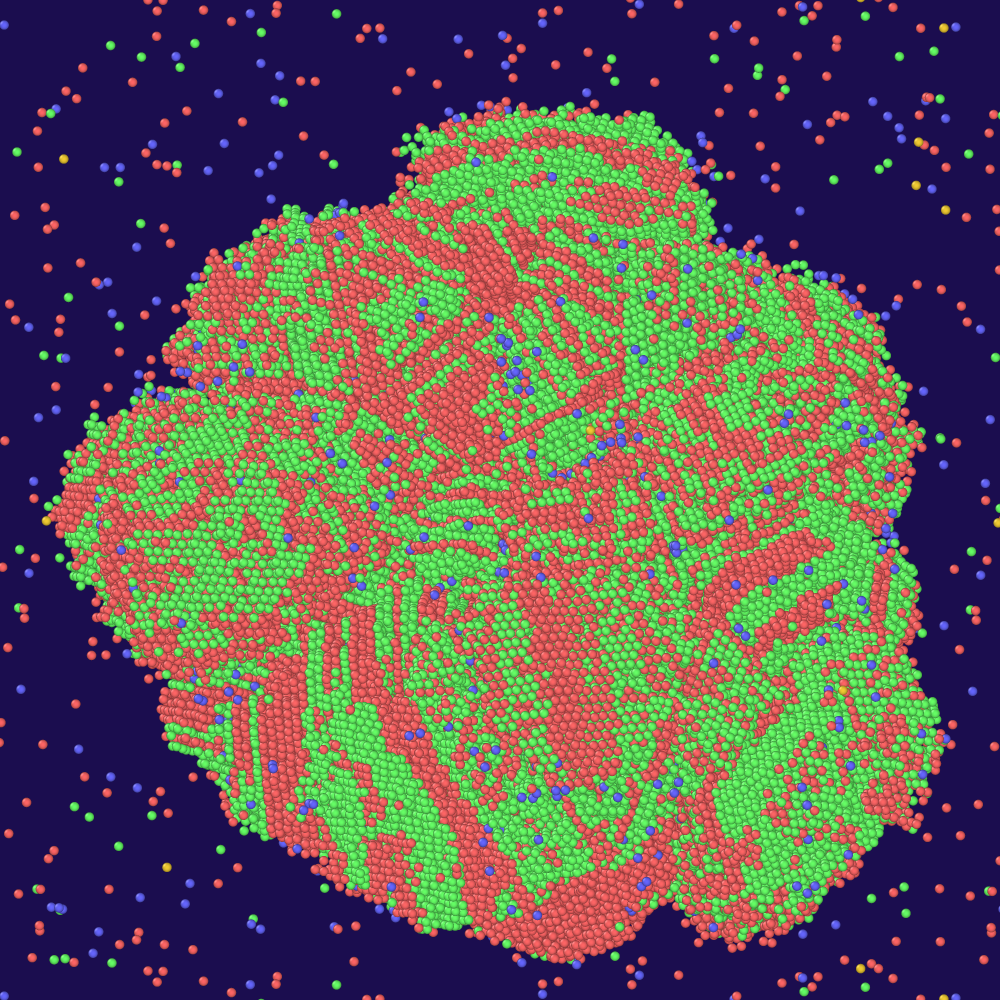}
    \includegraphics[width=0.3\linewidth]{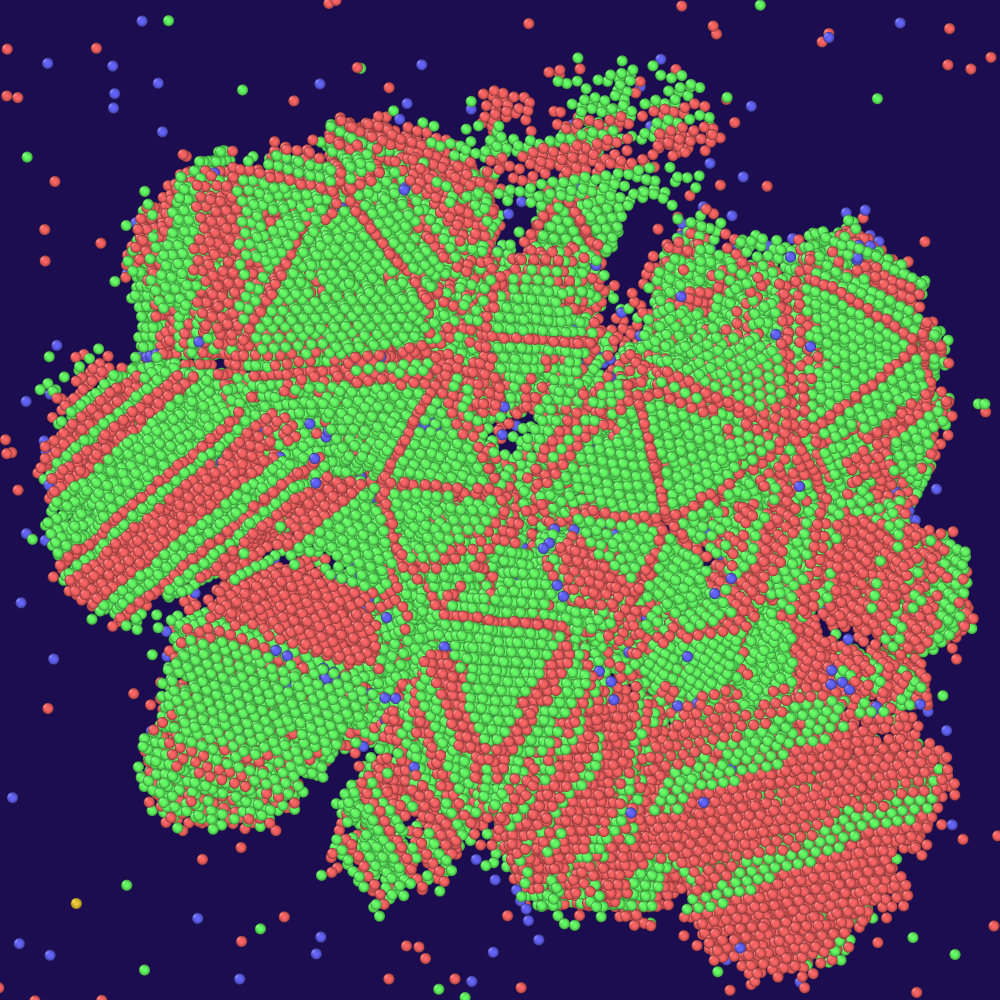}
\end{center}
\caption{Example of a cluster with icosahedral domains, $\eta=0.534$, Left: Cluster seen from the outside, right: Slice through the cluster. Red corresponds to hcp and green to fcc stacking, blue is bcc, and particles marked in orange have an icosahedral surrounding, particles in a liquid neighbourhood are not shown.}
\label{fig:snapshot2}
\end{figure}

\begin{figure}[h]
\begin{center}
    \includegraphics[width=0.3\linewidth]{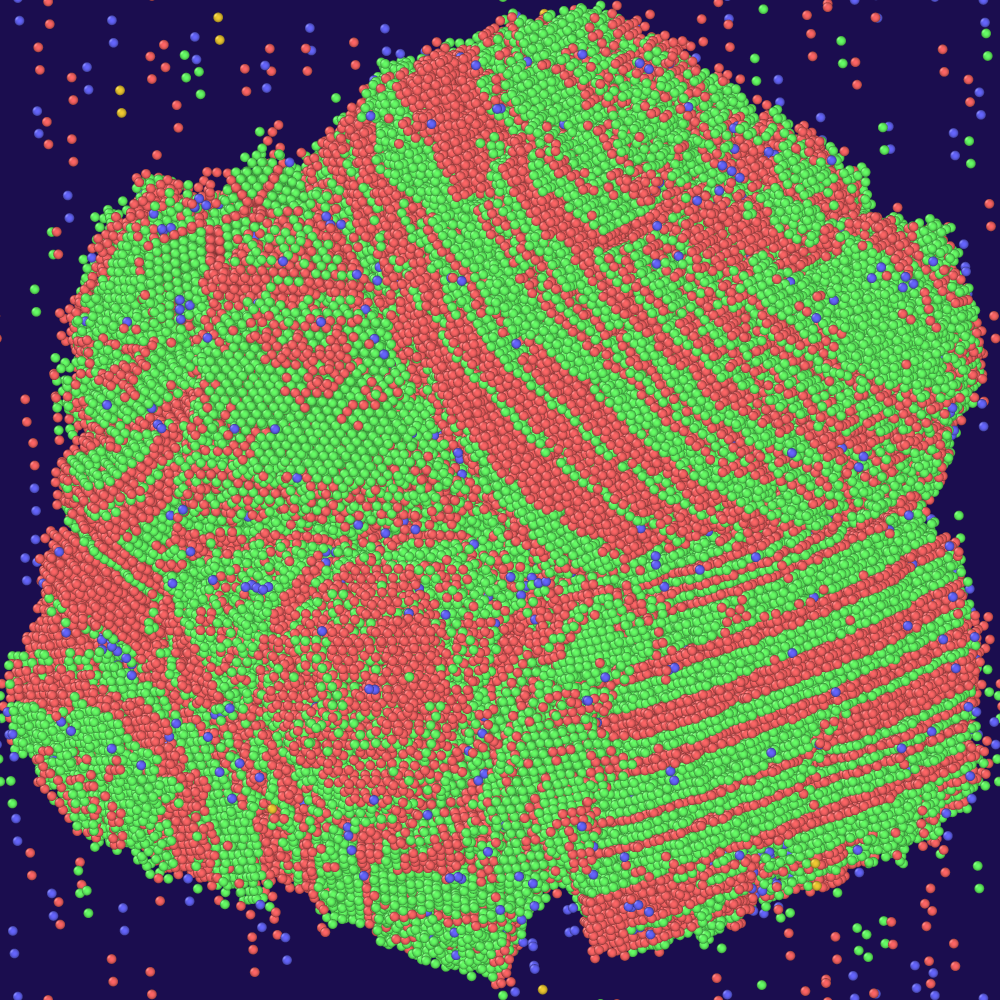}
    \includegraphics[width=0.3\linewidth]{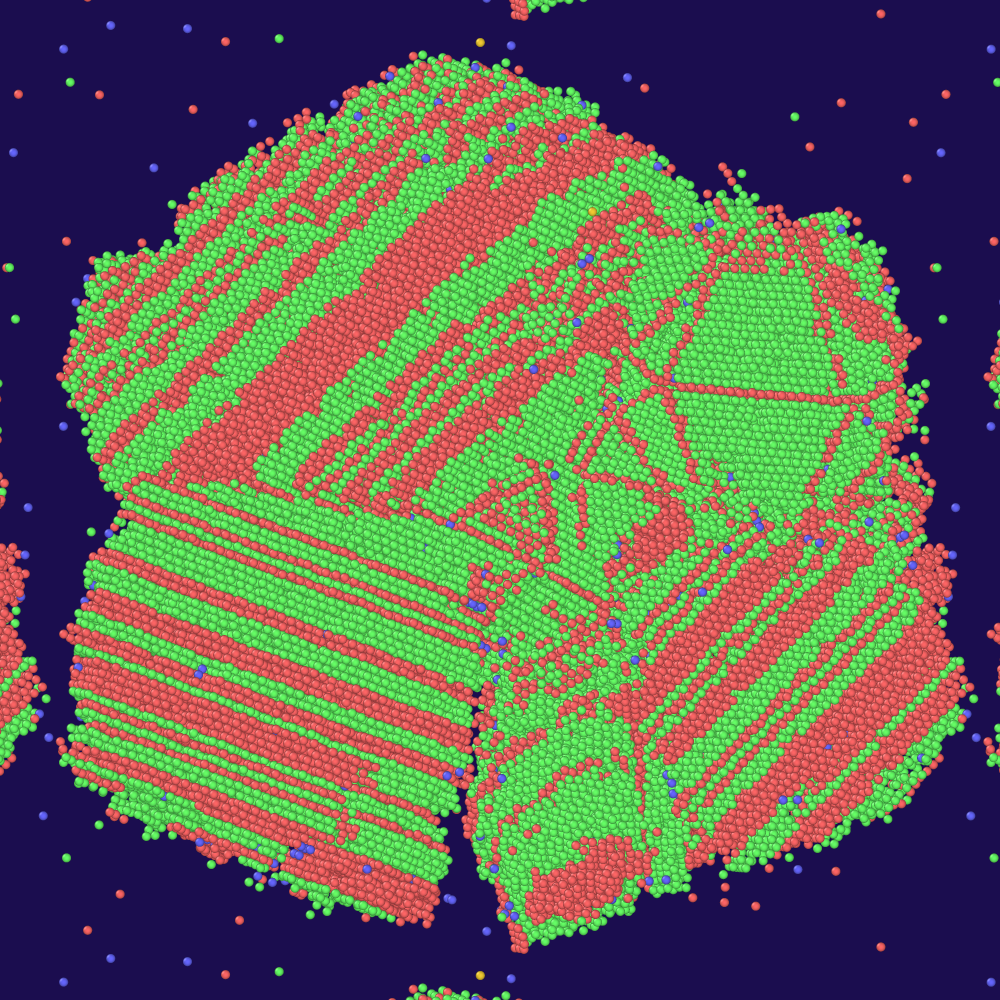}
\end{center}
\caption{Example of a cluster with a mixture of domains, $\eta=0.534$, Left: Cluster seen from the outside, right: Slice through the cluster. Red corresponds to hcp and green to fcc stacking, blue is bcc, and particles marked in orange have an icosahedral surrounding, particles in a liquid neighbourhood are not shown.}
\label{fig:snapshot3}
\end{figure}

\begin{figure}[h]
\begin{center}
    \includegraphics[width=0.3\linewidth]{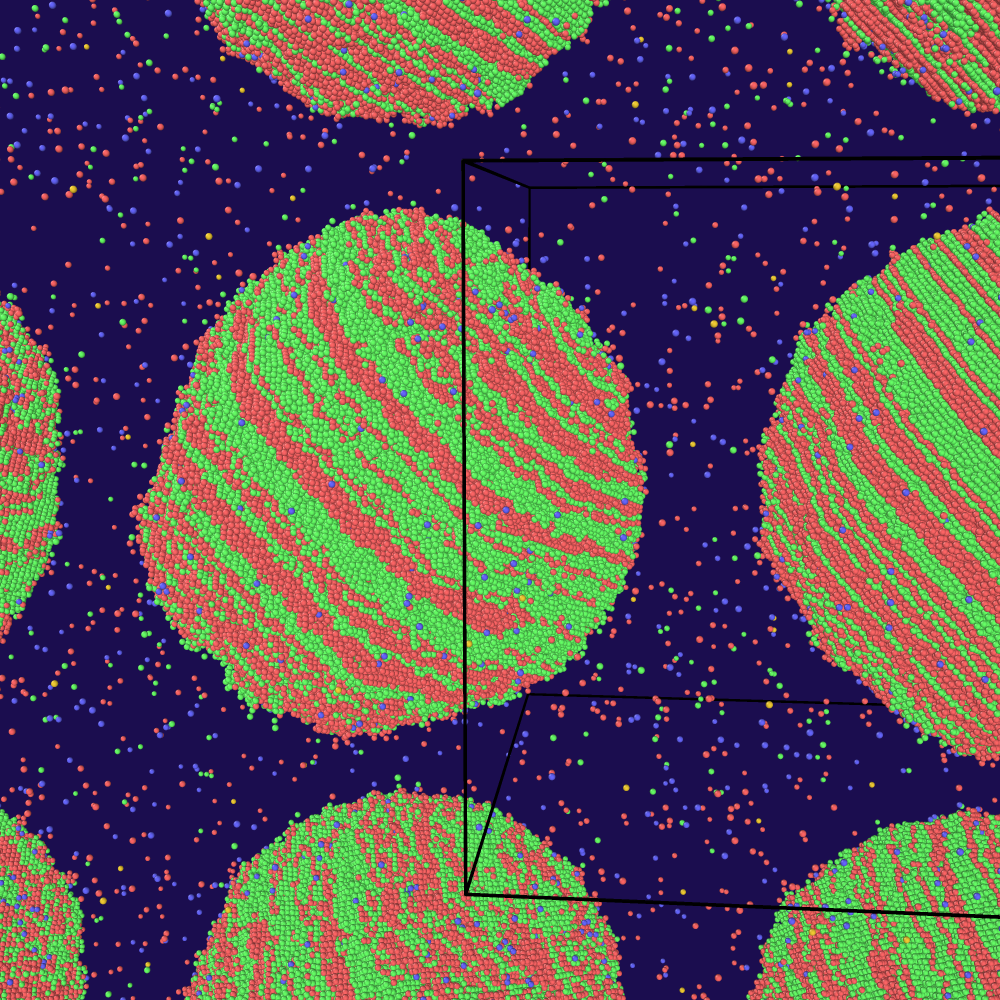}
    \includegraphics[width=0.3\linewidth]{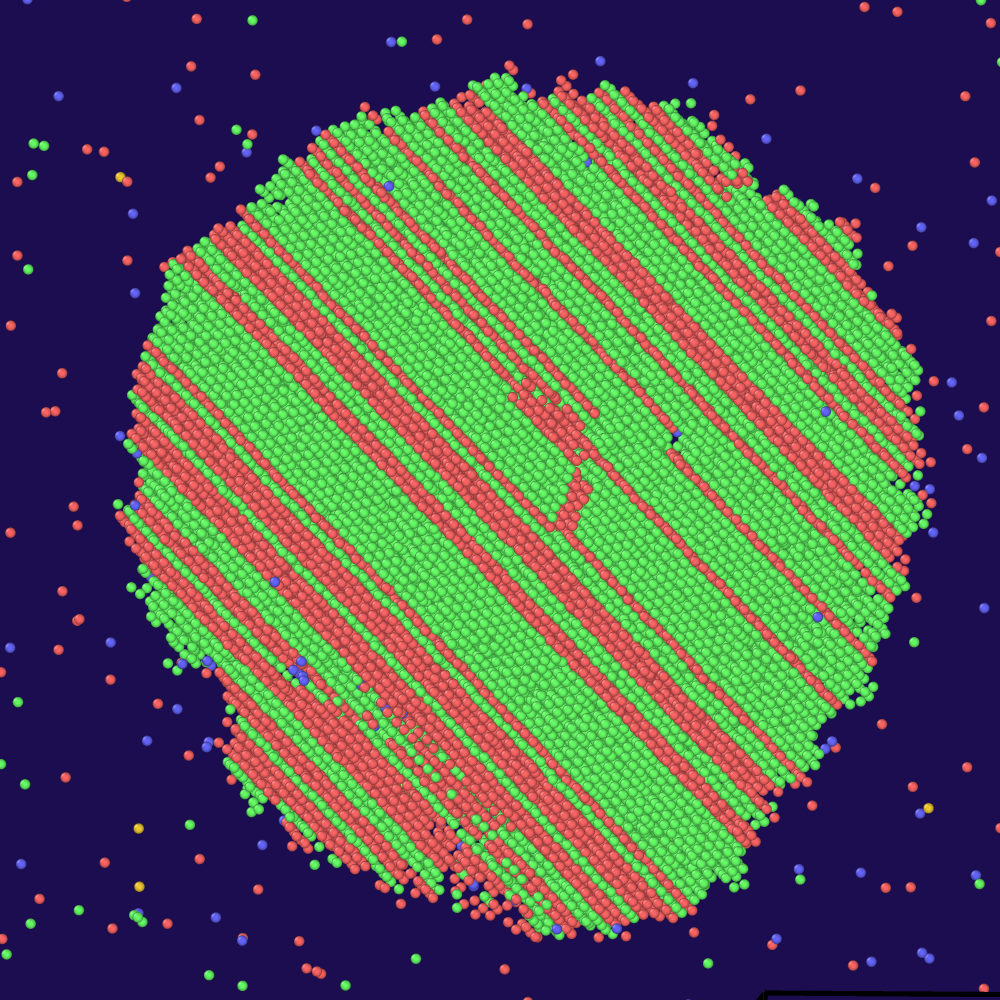}
\end{center}
\caption{Example of a mono-crystalline cluster, $\eta=0.534$, Left: Cluster seen from the outside, right: Slice through the cluster. Red corresponds to hcp and green to fcc stacking, blue is bcc, and particles marked in orange have an icosahedral surrounding, particles in a liquid neighbourhood are not shown. Repetitions are due to the periodic boundaries.}
\label{fig:snapshot4}
\end{figure}

\begin{figure}[h]
\begin{center}
    \includegraphics[width=0.3\linewidth]{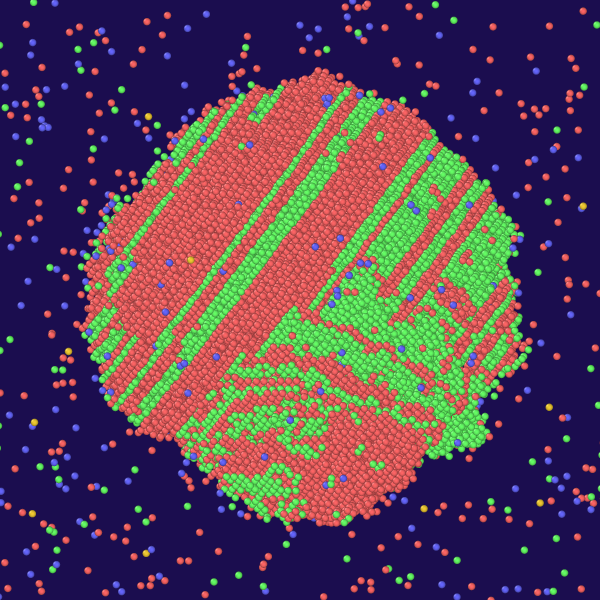}
    \includegraphics[width=0.3\linewidth]{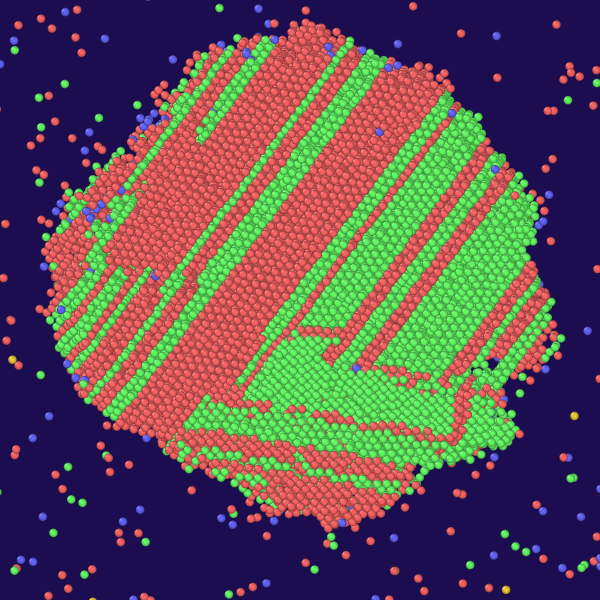}
\end{center}
\caption{Example of an fcc/hcp cluster, $\eta=0.531$, Left: Cluster seen from the outside, right: Slice through the cluster. Red corresponds to hcp and green to fcc stacking, blue is bcc, and particles marked in orange have an icosahedral surrounding, particles in a liquid neighbourhood are not shown.}
\label{fig:snapshot5}
\end{figure}

\begin{figure}[h]
\begin{center}
    \includegraphics[width=0.3\linewidth]{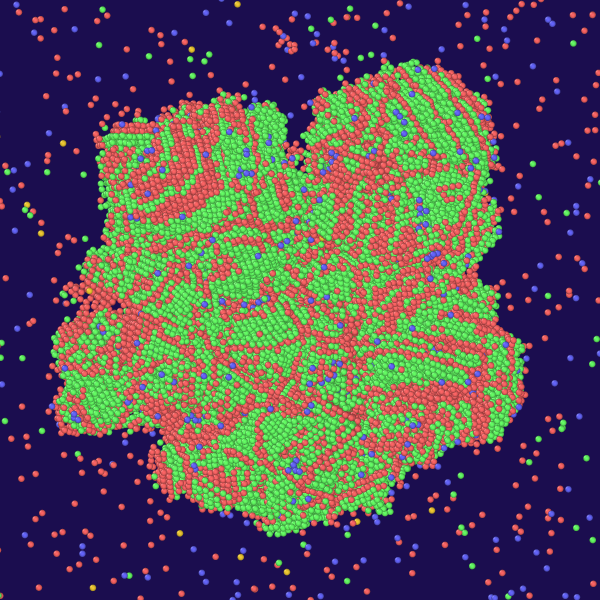}
    \includegraphics[width=0.3\linewidth]{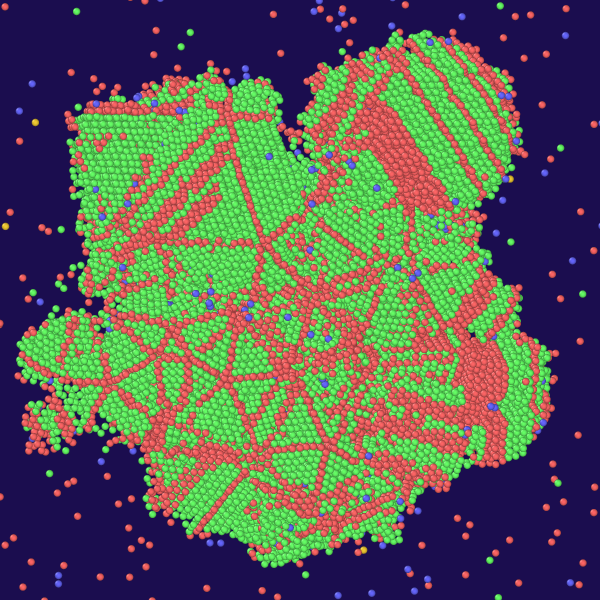}
\end{center}
\caption{Example of a mixed cluster containing many icosahedral domains, $\eta=0.531$, Left: Cluster seen from the outside, right: Slice through the cluster. Red corresponds to hcp and green to fcc stacking, blue is bcc, and particles marked in orange have an icosahedral surrounding, particles in a liquid neighbourhood are not shown.}
\label{fig:snapshot6}
\end{figure}

\begin{figure}[h]
\begin{center}
    \includegraphics[width=0.3\linewidth]{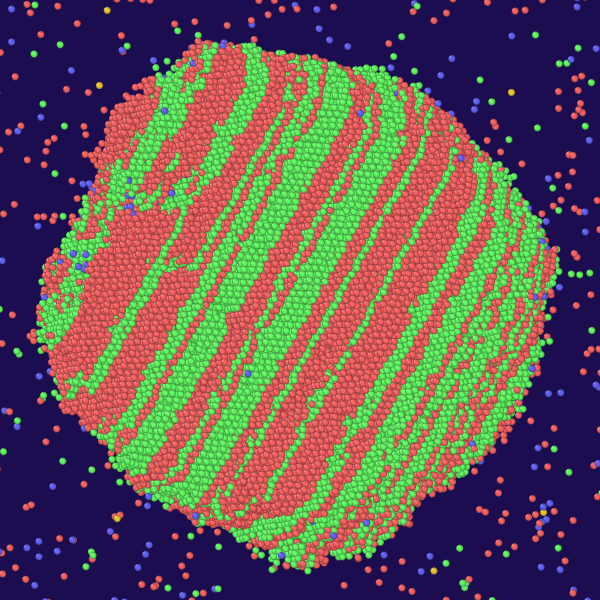}
    \includegraphics[width=0.3\linewidth]{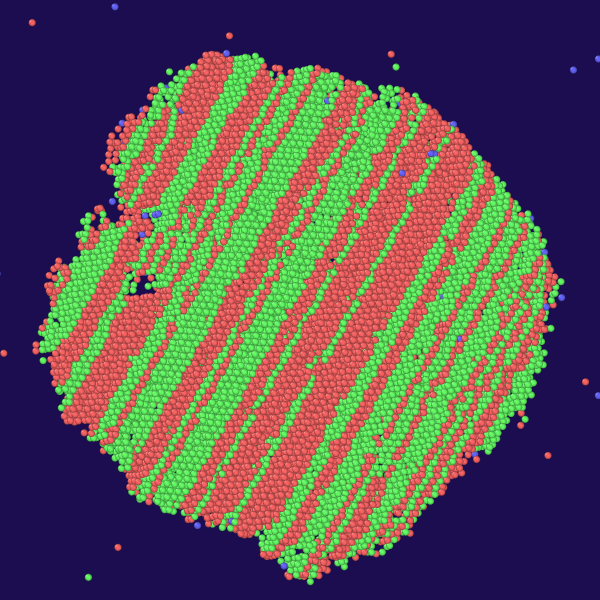}
\end{center}
\caption{Example of a mono-crystalline cluster, $\eta=0.531$, Left: Cluster seen from the outside, right: Slice through the cluster. Red corresponds to hcp and green to fcc stacking, blue is bcc, and particles marked in orange have an icosahedral surrounding, particles in a liquid neighbourhood are not shown.}
\label{fig:snapshot7}
\end{figure}

We used the common neighbour analysis tool included in the OVITO program \cite{ovito2010} to analyse simulation snapshots. We found fcc/hcp and icosahedral clusters, differing in domain numbers as can be seen in the snapshots Fig.~\ref{fig:snapshot1},\ref{fig:snapshot2},\ref{fig:snapshot3},\ref{fig:snapshot4} for volume fraction $\eta=0.534$ and in Fig.~\ref{fig:snapshot5},\ref{fig:snapshot6},\ref{fig:snapshot7} for $\eta=0.531$. The colouring is done by the OVITO common-neighbour-analysis tool. Red corresponds to hcp, green to fcc stacking, blue is bcc, and particles marked in orange have an icosahedral surrounding. The liquid is not shown.

The crystal in Fig.~\ref{fig:snapshot1} is clearly twinned. In Fig.~\ref{fig:snapshot2} and Fig.~\ref{fig:snapshot3}, there is a mixture of various twinned structures. Each cluster we observed had grown from one initial nucleus. We hardly ever observed the incorporation of an independently nucleated cluster into the growing cluster. Some domains formed by secondary nucleation on the cluster surface rather than by twinning, but this process was rare, too. Fig.~\ref{fig:snapshot4} shows a mono-crystalline cluster.

\newpage

\section{Rate of domain formation}
As shown in the example snapshots, the domains can easily be distinguished by eye (if only the order of magnitude of their rate of formation is of interest). While the fcc/hcp type clusters mostly contain fewer than ten domains, the icosahedral clusters typically incorporate more than ten domains for clusters with a few hundred thousand particles. We counted the overall number of domains in a cluster, $N_{\rm domain}$. With the corresponding number of particles in a cluster, $N_{\rm cluster}$, we calculated $\kappa_{\rm domain}$, the rate at which new domains formed on the surface of the cluster during the time of growth $T$ as follows:\\
\begin{align}
    N_{\rm domain} &= \int_0^T A_{\rm cluster}(t) \kappa_{\rm domain}(t)  dt\nonumber \\
    &= \int_0^T 4 \pi c^2 \kappa_{\rm domain} t^2 dt \nonumber\\
    &= \frac{4 \pi c^2}{3}  \kappa_{\rm domain} T^3\nonumber\\
    \label{eqn:kappa_domain_pre}
    \Leftrightarrow \kappa_{\rm domain} &= \frac{3 N_{\rm domain}}{4 \pi c^2 T^3} \quad .
\end{align}
Here, $A_{\rm cluster}(t)$ is the surface area of the cluster and $c$ is the radial growth rate.
We assumed that $\kappa_{\rm domain}$ is independent of time, which neglects the fact, that there might be an impact from the cluster's varying curvature on the twinning process. As we are not studying a large range of curvatures, this assumption is reasonable. Also, we assumed a constant spherical radial growth rate $c$, which is justified by our analysis presented below in the section regarding the crystal growth rate. The number of particles in the cluster $N_{\rm cluster}$ is linked to the cluster growth time $T$ and the radius of the cluster $R(t)$ by:
\begin{align}
    N_{\rm cluster}(t) &= \rho \frac{4 \pi }{3} R(t)^3\nonumber\\
    &=\rho \frac{4 \pi }{3} c^3 t^3\nonumber\\
    \label{eqn:T_N_dependence}
    \Leftrightarrow T^3 &= \frac{3 N_{\rm cluster}(T)}{4 \pi \rho c^3} \quad .
\end{align}
Inserting the relation given in eqn.~\ref{eqn:T_N_dependence} into eqn.~\ref{eqn:kappa_domain_pre} we obtain for $\kappa_{\rm domain}$
\begin{align}
    \label{eqn:kappa_domain_measurement}
    \kappa_{\rm domain} = \rho c \frac{N_{\rm domain}}{N_{\rm cluster}}\quad .
\end{align}

Regarding the dependence of $\kappa_{\rm domain}$ on the supersaturation of the liquid, we analysed the mono-disperse systems, ranging from $\eta = 0.531$ to $\eta = 0.54$. We could not find a definitive trend for $\kappa_{\rm domain}$. While this is partly due to our approximative method of determining $\kappa_{\rm domain}$, it is also reasonable to assume that $\kappa_{\rm domain}$ does not depend sensitively on the volume fraction of the system. Most of the domains are formed by successive twinning processes. The energy cost associated with the formation of a twinned domain stems from the cost of formation of a grain boundary, and the kinetics of twinning is driven by the elastic properties of the crystal. Both, the elastic constants as well as the cost of a grain boundary, depend only very weakly on the total volume fraction of the system. The dependence is due to the pressure applied on the crystal by the liquid, thus it is of a much weaker nature than e.g.~the dependence of the nucleation rate on the supersaturation.  We therefore carried out our analysis with a rate which is independent of the density of the system: $\kappa_{\rm domain} = 5 \cdot 10^{-4} (\sigma^2 \, \tau_{\rm L})^{-1}$.

\section{Crystal growth rate}
\label{sec:crystal_growth_rate}
We calculated shape descriptors based on the tensor of gyration of the crystallites and found the crystallites to be mostly spherical for particle numbers above a few thousand. Furthermore, we calculated power law fits to the particle number growth law of the largest cluster within the system. A quantitative evaluation of the power law coefficient, $n$, for the 5\% Gaussian poly-disperse system at $\eta = 55.4\%$, resulted in $n = 2.97 \pm 0.03$, consistent with the theoretical value for constant radial growth being $n = 3$. 

Qualitative deviations from this behaviour were not observed in any system, except for fluctuations at cluster sizes below 1000 particles. Therefore, we concluded that constant radial growth is the dominant growth mode in the system.

The cluster growth rates were measured by linear regressions to the third root of the particle number in the largest cluster. In Fig.~\ref{fig:growth} we show the data of a poly-disperse seeded system as an example.

While we are confident that our measurement is accurate, we also note, that the radial growth rate enters the calculation of the crystallization rate only as a fourth root. Thus, to shift the rate densities presented in the letter by one order of magnitude, the value of $c$ would have to vary by a factor of 10,000. 

\begin{figure}
\label{fig:growth}
	\includegraphics[width = 0.6\linewidth]{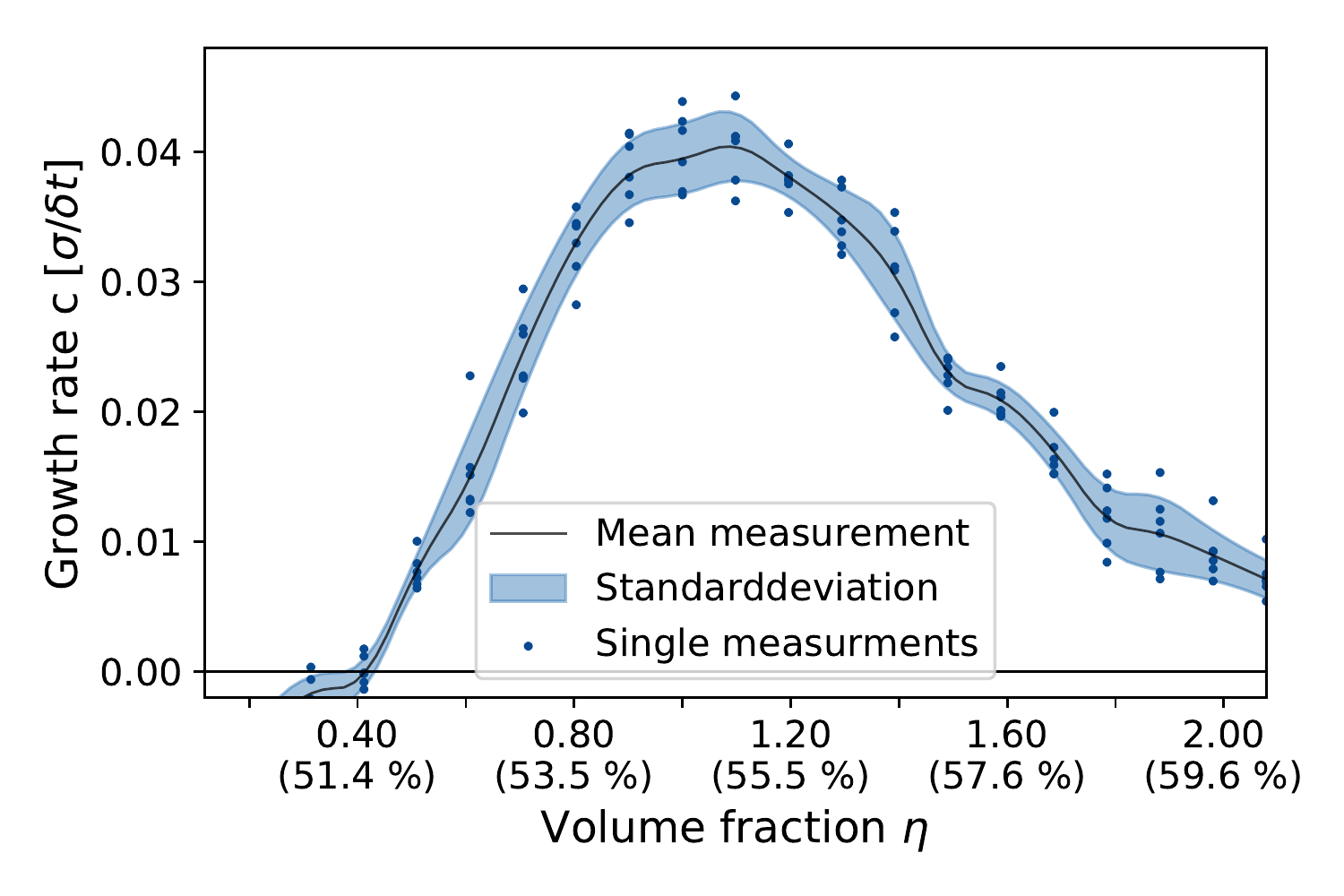}
	\caption{
		Cluster growth rate, measured in a 5\% Gaussian poly-disperse system of \SI{500000}{\text{particles}}, with an initial seed of about \SI{7000}{\text{particles}}. 
	}
	\label{fig:growth}
\end{figure}

\section{JMAK model}
In the main text we used the expression
\begin{align}
\label{eqn:Avrami_crystallinity}
X_s(t)= 1 - \exp \left( - \frac{\pi}{3} c^3 \kappa_{\text{hom}} t^4 \right) 
\end{align}
to model how the fraction of crystalline material in a system depends on time, if crystalline nuclei form at a rate $\kappa_{\text{hom}}$ and grow spherically with a radial growth rate $c$. The reasoning that leads to this equation is the following:

The number of crystalline clusters formed in a time interval of length ${\rm d}t$ in a system of volume $V$ is
\[
\frac{{\rm d}N_{\rm{cluster}}}{{\rm d} t} = V\kappa_{\text{hom}}  \quad .
\]
At time $t$, a cluster which was formed at a time $t'$ has grown to a volume 
\[
V_{\rm{cluster}}(t,t') = \frac{4\pi}{3}c^3(t-t')^3 \quad .
\]

Thus in the "ideal case", in which the clusters do not overlap, 
the total volume filled by crystals at time $t$, $V_{\rm{cryst,\; ideal}}(t)$, would be given by
\[
V_{\rm{cryst,\; ideal}}(t) = \int_0^t \frac{{\rm d}N_{\rm{cluster}}}{{\rm d} t} V_{\rm{cluster}}(t,t'){\rm d}t' = \int_0^t V\kappa_{\rm hom} \frac{4\pi}{3}c^3 (t-t')^3 {\rm d}t' \quad ,
\]
i.e.~per infinitesimal time interval, the volume filled by crystals grows as
\[
{\rm d}V_{\rm{cryst,\; ideal}} = V\kappa_{\rm hom}  \frac{4 \pi}{3}c^3 t^3 {\rm d}t \quad .
\]

However, in a real system, the volume available for nucleation and growth shrinks while the crystalline clusters grow. Therefore all processes need to be  weighted with the probability of still finding non-transformed space. If the actually crystallized volume is $V_{\rm cryst}(t)$, then the remaining volume fraction is given by $1-\frac{V_{\rm  cryst}(t)}{V}$. Expressed in the differential form:
\[
{\rm d}V_{\rm cryst} = \left(1-\frac{V_{ \rm cryst}}{V} \right) {\rm d} V_{\rm{\rm cryst, ideal}} = \left(1-\frac{V_{\rm cryst}}{V} \right)  V \kappa_{\text{hom}}  \frac{ 4 \pi}{3}c^3 t^3 {\rm d}t \quad .
\]
Using the definition of the crystalline volume fraction, $X_s := \frac{V_{\rm cryst}}{V}$, we find
\[
V {\rm d}X_s = \left(1-X_s \right)  V \kappa_{\rm hom}  \frac{4 \pi}{3}c^3 t^3 {\rm d}t \quad
\]
\[
\Leftrightarrow \left(1-X_s \right)^{-1}   {\rm d}X_s = \kappa_{\rm hom}  \frac{4 \pi}{3}c^3 t^3 {\rm d}t \quad
\]
\[
\Leftrightarrow - \log \left(1-X_s \right) = \kappa_{\rm hom}  \frac{\pi}{3}c^3 t^4 \quad
\]
\[
\Leftrightarrow  X_s(t) = 1 - \exp \left( -\kappa_{\rm hom}  \frac{\pi}{3}c^3 t^4 \right) \quad .
\]

The final result corresponds to the expression given in eq.~\ref{eqn:Avrami_crystallinity}, which is used in the article.


\begin{thebibliography}{41}%
\makeatletter
\providecommand \@ifxundefined [1]{%
 \@ifx{#1\undefined}
}%
\providecommand \@ifnum [1]{%
 \ifnum #1\expandafter \@firstoftwo
 \else \expandafter \@secondoftwo
 \fi
}%
\providecommand \@ifx [1]{%
 \ifx #1\expandafter \@firstoftwo
 \else \expandafter \@secondoftwo
 \fi
}%
\providecommand \natexlab [1]{#1}%
\providecommand \enquote  [1]{``#1''}%
\providecommand \bibnamefont  [1]{#1}%
\providecommand \bibfnamefont [1]{#1}%
\providecommand \citenamefont [1]{#1}%
\providecommand \href@noop [0]{\@secondoftwo}%
\providecommand \href [0]{\begingroup \@sanitize@url \@href}%
\providecommand \@href[1]{\@@startlink{#1}\@@href}%
\providecommand \@@href[1]{\endgroup#1\@@endlink}%
\providecommand \@sanitize@url [0]{\catcode `\\12\catcode `\$12\catcode
  `\&12\catcode `\#12\catcode `\^12\catcode `\_12\catcode `\%12\relax}%
\providecommand \@@startlink[1]{}%
\providecommand \@@endlink[0]{}%
\providecommand \url  [0]{\begingroup\@sanitize@url \@url }%
\providecommand \@url [1]{\endgroup\@href {#1}{\urlprefix }}%
\providecommand \urlprefix  [0]{URL }%
\providecommand \Eprint [0]{\href }%
\providecommand \doibase [0]{http://dx.doi.org/}%
\providecommand \selectlanguage [0]{\@gobble}%
\providecommand \bibinfo  [0]{\@secondoftwo}%
\providecommand \bibfield  [0]{\@secondoftwo}%
\providecommand \translation [1]{[#1]}%
\providecommand \BibitemOpen [0]{}%
\providecommand \bibitemStop [0]{}%
\providecommand \bibitemNoStop [0]{.\EOS\space}%
\providecommand \EOS [0]{\spacefactor3000\relax}%
\providecommand \BibitemShut  [1]{\csname bibitem#1\endcsname}%
\let\auto@bib@innerbib\@empty
\bibitem [{\citenamefont {Alder}\ and\ \citenamefont
  {Wainwright}(1959)}]{Alders59}%
  \BibitemOpen
  \bibfield  {author} {\bibinfo {author} {\bibfnamefont {B.~J.}\ \bibnamefont
  {Alder}}\ and\ \bibinfo {author} {\bibfnamefont {T.~E.}\ \bibnamefont
  {Wainwright}},\ }\href {\doibase 10.1063/1.1730376} {\bibfield  {journal}
  {\bibinfo  {journal} {The Journal of Chemical Physics}\ }\textbf {\bibinfo
  {volume} {31}},\ \bibinfo {pages} {459} (\bibinfo {year} {1959})}\BibitemShut
  {NoStop}%
\bibitem [{\citenamefont {Pusey}\ and\ \citenamefont {van
  Megen}(1986)}]{Pusey1986}%
  \BibitemOpen
  \bibfield  {author} {\bibinfo {author} {\bibfnamefont {P.~N.}\ \bibnamefont
  {Pusey}}\ and\ \bibinfo {author} {\bibfnamefont {W.}~\bibnamefont {van
  Megen}},\ }\href {https://www.nature.com/articles/320340a0} {\bibfield
  {journal} {\bibinfo  {journal} {Nature}\ }\textbf {\bibinfo {volume} {320}},\
  \bibinfo {pages} {340} (\bibinfo {year} {1986})}\BibitemShut {NoStop}%
\bibitem [{\citenamefont {Auer}\ and\ \citenamefont
  {Frenkel}(2001)}]{Auer2001}%
  \BibitemOpen
  \bibfield  {author} {\bibinfo {author} {\bibfnamefont {S.}~\bibnamefont
  {Auer}}\ and\ \bibinfo {author} {\bibfnamefont {D.}~\bibnamefont {Frenkel}},\
  }\href {\doibase 10.1038/35059035} {\bibfield  {journal} {\bibinfo  {journal}
  {Nature}\ }\textbf {\bibinfo {volume} {409}},\ \bibinfo {pages} {1020}
  (\bibinfo {year} {2001})}\BibitemShut {NoStop}%
\bibitem [{\citenamefont {Filion}\ \emph
  {et~al.}(2010{\natexlab{a}})\citenamefont {Filion}, \citenamefont {Hermes},
  \citenamefont {Ni},\ and\ \citenamefont {Dijkstra}}]{Filion2010}%
  \BibitemOpen
  \bibfield  {author} {\bibinfo {author} {\bibfnamefont {L.}~\bibnamefont
  {Filion}}, \bibinfo {author} {\bibfnamefont {M.}~\bibnamefont {Hermes}},
  \bibinfo {author} {\bibfnamefont {R.}~\bibnamefont {Ni}}, \ and\ \bibinfo
  {author} {\bibfnamefont {M.}~\bibnamefont {Dijkstra}},\ }\href {\doibase
  10.1063/1.3506838} {\bibfield  {journal} {\bibinfo  {journal} {Journal of
  Chemical Physics}\ }\textbf {\bibinfo {volume} {133}} (\bibinfo {year}
  {2010}{\natexlab{a}}),\ 10.1063/1.3506838}\BibitemShut {NoStop}%
\bibitem [{\citenamefont {Filion}\ \emph {et~al.}(2011)\citenamefont {Filion},
  \citenamefont {Ni}, \citenamefont {Frenkel},\ and\ \citenamefont
  {Dijkstra}}]{Filion2011}%
  \BibitemOpen
  \bibfield  {author} {\bibinfo {author} {\bibfnamefont {L.}~\bibnamefont
  {Filion}}, \bibinfo {author} {\bibfnamefont {R.}~\bibnamefont {Ni}}, \bibinfo
  {author} {\bibfnamefont {D.}~\bibnamefont {Frenkel}}, \ and\ \bibinfo
  {author} {\bibfnamefont {M.}~\bibnamefont {Dijkstra}},\ }\href {\doibase
  10.1063/1.3572059} {\bibfield  {journal} {\bibinfo  {journal} {Journal of
  Chemical Physics}\ }\textbf {\bibinfo {volume} {134}} (\bibinfo {year}
  {2011}),\ 10.1063/1.3572059}\BibitemShut {NoStop}%
\bibitem [{\citenamefont {Russo}\ \emph {et~al.}(2013)\citenamefont {Russo},
  \citenamefont {Maggs}, \citenamefont {Bonn},\ and\ \citenamefont
  {Tanaka}}]{russo2013}%
  \BibitemOpen
  \bibfield  {author} {\bibinfo {author} {\bibfnamefont {J.}~\bibnamefont
  {Russo}}, \bibinfo {author} {\bibfnamefont {A.~C.}\ \bibnamefont {Maggs}},
  \bibinfo {author} {\bibfnamefont {D.}~\bibnamefont {Bonn}}, \ and\ \bibinfo
  {author} {\bibfnamefont {H.}~\bibnamefont {Tanaka}},\ }\href@noop {}
  {\bibfield  {journal} {\bibinfo  {journal} {Soft Matter}\ }\textbf {\bibinfo
  {volume} {9}},\ \bibinfo {pages} {7369} (\bibinfo {year} {2013})}\BibitemShut
  {NoStop}%
\bibitem [{\citenamefont {Russo}\ and\ \citenamefont
  {Tanaka}(2016)}]{russo2016}%
  \BibitemOpen
  \bibfield  {author} {\bibinfo {author} {\bibfnamefont {J.}~\bibnamefont
  {Russo}}\ and\ \bibinfo {author} {\bibfnamefont {H.}~\bibnamefont {Tanaka}},\
  }\href@noop {} {\bibfield  {journal} {\bibinfo  {journal} {MRS Bulletin}\
  }\textbf {\bibinfo {volume} {41}},\ \bibinfo {pages} {369} (\bibinfo {year}
  {2016})}\BibitemShut {NoStop}%
\bibitem [{\citenamefont {Wood}\ \emph {et~al.}(2018)\citenamefont {Wood},
  \citenamefont {Russo}, \citenamefont {Turci},\ and\ \citenamefont
  {Royall}}]{wood2018}%
  \BibitemOpen
  \bibfield  {author} {\bibinfo {author} {\bibfnamefont {N.}~\bibnamefont
  {Wood}}, \bibinfo {author} {\bibfnamefont {J.}~\bibnamefont {Russo}},
  \bibinfo {author} {\bibfnamefont {F.}~\bibnamefont {Turci}}, \ and\ \bibinfo
  {author} {\bibfnamefont {C.~P.}\ \bibnamefont {Royall}},\ }\href@noop {}
  {\bibfield  {journal} {\bibinfo  {journal} {The Journal of chemical physics}\
  }\textbf {\bibinfo {volume} {149}},\ \bibinfo {pages} {204506} (\bibinfo
  {year} {2018})}\BibitemShut {NoStop}%
\bibitem [{\citenamefont {Auer}\ and\ \citenamefont
  {Frenkel}(2002)}]{auer2002}%
  \BibitemOpen
  \bibfield  {author} {\bibinfo {author} {\bibfnamefont {S.}~\bibnamefont
  {Auer}}\ and\ \bibinfo {author} {\bibfnamefont {D.}~\bibnamefont {Frenkel}},\
  }\href@noop {} {\bibfield  {journal} {\bibinfo  {journal} {Journal of
  Physics: Condensed Matter}\ }\textbf {\bibinfo {volume} {14}},\ \bibinfo
  {pages} {7667} (\bibinfo {year} {2002})}\BibitemShut {NoStop}%
\bibitem [{\citenamefont {Radu}\ and\ \citenamefont
  {Schilling}(2014)}]{Radu2014}%
  \BibitemOpen
  \bibfield  {author} {\bibinfo {author} {\bibfnamefont {M.}~\bibnamefont
  {Radu}}\ and\ \bibinfo {author} {\bibfnamefont {T.}~\bibnamefont
  {Schilling}},\ }\href {\doibase 10.1209/0295-5075/105/26001} {\bibfield
  {journal} {\bibinfo  {journal} {Epl}\ }\textbf {\bibinfo {volume} {105}},\
  \bibinfo {pages} {1} (\bibinfo {year} {2014})}\BibitemShut {NoStop}%
\bibitem [{\citenamefont {Tateno}\ \emph {et~al.}(2019)\citenamefont {Tateno},
  \citenamefont {Yanagishima}, \citenamefont {Russo},\ and\ \citenamefont
  {Tanaka}}]{Tateno2019}%
  \BibitemOpen
  \bibfield  {author} {\bibinfo {author} {\bibfnamefont {M.}~\bibnamefont
  {Tateno}}, \bibinfo {author} {\bibfnamefont {T.}~\bibnamefont {Yanagishima}},
  \bibinfo {author} {\bibfnamefont {J.}~\bibnamefont {Russo}}, \ and\ \bibinfo
  {author} {\bibfnamefont {H.}~\bibnamefont {Tanaka}},\ }\href {\doibase
  10.1103/PhysRevLett.123.258002} {\bibfield  {journal} {\bibinfo  {journal}
  {Physical Review Letters}\ }\textbf {\bibinfo {volume} {123}},\ \bibinfo
  {pages} {258002} (\bibinfo {year} {2019})}\BibitemShut {NoStop}%
\bibitem [{\citenamefont {Fiorucci}\ \emph
  {et~al.}(2020{\natexlab{a}})\citenamefont {Fiorucci}, \citenamefont {Coli},
  \citenamefont {Padding},\ and\ \citenamefont {Dijkstra}}]{Fiorucci2020}%
  \BibitemOpen
  \bibfield  {author} {\bibinfo {author} {\bibfnamefont {G.}~\bibnamefont
  {Fiorucci}}, \bibinfo {author} {\bibfnamefont {G.~M.}\ \bibnamefont {Coli}},
  \bibinfo {author} {\bibfnamefont {J.~T.}\ \bibnamefont {Padding}}, \ and\
  \bibinfo {author} {\bibfnamefont {M.}~\bibnamefont {Dijkstra}},\ }\href
  {\doibase 10.1063/1.5137815} {\bibfield  {journal} {\bibinfo  {journal}
  {Journal of Chemical Physics}\ }\textbf {\bibinfo {volume} {152}},\ \bibinfo
  {pages} {064903} (\bibinfo {year} {2020}{\natexlab{a}})}\BibitemShut
  {NoStop}%
\bibitem [{\citenamefont {Pusey}\ \emph {et~al.}(2009)\citenamefont {Pusey},
  \citenamefont {Zaccarelli}, \citenamefont {Valeriani}, \citenamefont {Sanz},
  \citenamefont {Poon},\ and\ \citenamefont {Cates}}]{Pusey2009}%
  \BibitemOpen
  \bibfield  {author} {\bibinfo {author} {\bibfnamefont {P.~N.}\ \bibnamefont
  {Pusey}}, \bibinfo {author} {\bibfnamefont {E.}~\bibnamefont {Zaccarelli}},
  \bibinfo {author} {\bibfnamefont {C.}~\bibnamefont {Valeriani}}, \bibinfo
  {author} {\bibfnamefont {E.}~\bibnamefont {Sanz}}, \bibinfo {author}
  {\bibfnamefont {W.~C.}\ \bibnamefont {Poon}}, \ and\ \bibinfo {author}
  {\bibfnamefont {M.~E.}\ \bibnamefont {Cates}},\ }\href {\doibase
  10.1098/rsta.2009.0181} {\bibfield  {journal} {\bibinfo  {journal}
  {Philosophical Transactions of the Royal Society A: Mathematical, Physical
  and Engineering Sciences}\ }\textbf {\bibinfo {volume} {367}},\ \bibinfo
  {pages} {4993} (\bibinfo {year} {2009})}\BibitemShut {NoStop}%
\bibitem [{\citenamefont {Poon}\ \emph {et~al.}(2012)\citenamefont {Poon},
  \citenamefont {Weeks},\ and\ \citenamefont {Royall}}]{Poon2012}%
  \BibitemOpen
  \bibfield  {author} {\bibinfo {author} {\bibfnamefont {W.~C.}\ \bibnamefont
  {Poon}}, \bibinfo {author} {\bibfnamefont {E.~R.}\ \bibnamefont {Weeks}}, \
  and\ \bibinfo {author} {\bibfnamefont {C.~P.}\ \bibnamefont {Royall}},\
  }\href@noop {} {\bibfield  {journal} {\bibinfo  {journal} {Soft Matter}\
  }\textbf {\bibinfo {volume} {8}},\ \bibinfo {pages} {21} (\bibinfo {year}
  {2012})}\BibitemShut {NoStop}%
\bibitem [{\citenamefont {Royall}\ \emph {et~al.}(2013)\citenamefont {Royall},
  \citenamefont {Poon},\ and\ \citenamefont {Weeks}}]{Royall2013}%
  \BibitemOpen
  \bibfield  {author} {\bibinfo {author} {\bibfnamefont {C.~P.}\ \bibnamefont
  {Royall}}, \bibinfo {author} {\bibfnamefont {W.~C.}\ \bibnamefont {Poon}}, \
  and\ \bibinfo {author} {\bibfnamefont {E.~R.}\ \bibnamefont {Weeks}},\
  }\href@noop {} {\bibfield  {journal} {\bibinfo  {journal} {Soft Matter}\
  }\textbf {\bibinfo {volume} {9}},\ \bibinfo {pages} {17} (\bibinfo {year}
  {2013})}\BibitemShut {NoStop}%
\bibitem [{\citenamefont {Sear}(2012)}]{Sear2012}%
  \BibitemOpen
  \bibfield  {author} {\bibinfo {author} {\bibfnamefont {R.~P.}\ \bibnamefont
  {Sear}},\ }\href {\doibase 10.1179/1743280411Y.0000000015} {\bibfield
  {journal} {\bibinfo  {journal} {International Materials Reviews}\ }\textbf
  {\bibinfo {volume} {57}},\ \bibinfo {pages} {328} (\bibinfo {year}
  {2012})}\BibitemShut {NoStop}%
\bibitem [{\citenamefont {Schilling}\ \emph {et~al.}(2011)\citenamefont
  {Schilling}, \citenamefont {Dorosz}, \citenamefont {Sch{\"{o}}pe},\ and\
  \citenamefont {Opletal}}]{Schilling2011}%
  \BibitemOpen
  \bibfield  {author} {\bibinfo {author} {\bibfnamefont {T.}~\bibnamefont
  {Schilling}}, \bibinfo {author} {\bibfnamefont {S.}~\bibnamefont {Dorosz}},
  \bibinfo {author} {\bibfnamefont {H.~J.}\ \bibnamefont {Sch{\"{o}}pe}}, \
  and\ \bibinfo {author} {\bibfnamefont {G.}~\bibnamefont {Opletal}},\ }\href
  {\doibase 10.1088/0953-8984/23/19/194120} {\bibfield  {journal} {\bibinfo
  {journal} {Journal of Physics Condensed Matter}\ }\textbf {\bibinfo {volume}
  {23}},\ \bibinfo {pages} {194120} (\bibinfo {year} {2011})}\BibitemShut
  {NoStop}%
\bibitem [{\citenamefont {Kuhnhold}\ \emph {et~al.}(2019)\citenamefont
  {Kuhnhold}, \citenamefont {Meyer}, \citenamefont {Amati}, \citenamefont
  {Philipp},\ and\ \citenamefont {Schilling}}]{Kuhnhold2019}%
  \BibitemOpen
  \bibfield  {author} {\bibinfo {author} {\bibfnamefont {A.}~\bibnamefont
  {Kuhnhold}}, \bibinfo {author} {\bibfnamefont {H.}~\bibnamefont {Meyer}},
  \bibinfo {author} {\bibfnamefont {G.}~\bibnamefont {Amati}}, \bibinfo
  {author} {\bibfnamefont {P.}~\bibnamefont {Philipp}}, \ and\ \bibinfo
  {author} {\bibfnamefont {T.}~\bibnamefont {Schilling}},\ }\href {\doibase
  10.1103/PhysRevE.100.052140} {\bibfield  {journal} {\bibinfo  {journal}
  {Physical Review E}\ }\textbf {\bibinfo {volume} {100}},\ \bibinfo {pages}
  {52140} (\bibinfo {year} {2019})}\BibitemShut {NoStop}%
\bibitem [{\citenamefont {Meyer}\ \emph {et~al.}(2021)\citenamefont {Meyer},
  \citenamefont {Glatzel}, \citenamefont {W{\"{o}}hler},\ and\ \citenamefont
  {Schilling}}]{Meyer2021}%
  \BibitemOpen
  \bibfield  {author} {\bibinfo {author} {\bibfnamefont {H.}~\bibnamefont
  {Meyer}}, \bibinfo {author} {\bibfnamefont {F.}~\bibnamefont {Glatzel}},
  \bibinfo {author} {\bibfnamefont {W.}~\bibnamefont {W{\"{o}}hler}}, \ and\
  \bibinfo {author} {\bibfnamefont {T.}~\bibnamefont {Schilling}},\ }\href
  {\doibase 10.1103/PhysRevE.103.022102} {\bibfield  {journal} {\bibinfo
  {journal} {Physical Review E}\ }\textbf {\bibinfo {volume} {103}},\ \bibinfo
  {pages} {22102} (\bibinfo {year} {2021})}\BibitemShut {NoStop}%
\bibitem [{\citenamefont {Harland}\ and\ \citenamefont {van
  Megen}(1997)}]{Harland1997}%
  \BibitemOpen
  \bibfield  {author} {\bibinfo {author} {\bibfnamefont {J.~L.}\ \bibnamefont
  {Harland}}\ and\ \bibinfo {author} {\bibfnamefont {W.}~\bibnamefont {van
  Megen}},\ }\href {\doibase 10.1103/PhysRevE.55.3054} {\bibfield  {journal}
  {\bibinfo  {journal} {Physical Review E}\ }\textbf {\bibinfo {volume} {55}},\
  \bibinfo {pages} {3054} (\bibinfo {year} {1997})}\BibitemShut {NoStop}%
\bibitem [{\citenamefont {He}\ \emph {et~al.}(1996)\citenamefont {He},
  \citenamefont {Ackerson}, \citenamefont {van Megen}, \citenamefont
  {Underwood},\ and\ \citenamefont {Sch{\"{a}}tzel}}]{He1996}%
  \BibitemOpen
  \bibfield  {author} {\bibinfo {author} {\bibfnamefont {Y.}~\bibnamefont
  {He}}, \bibinfo {author} {\bibfnamefont {B.~J.}\ \bibnamefont {Ackerson}},
  \bibinfo {author} {\bibfnamefont {W.}~\bibnamefont {van Megen}}, \bibinfo
  {author} {\bibfnamefont {S.~M.}\ \bibnamefont {Underwood}}, \ and\ \bibinfo
  {author} {\bibfnamefont {K.}~\bibnamefont {Sch{\"{a}}tzel}},\ }\href
  {\doibase 10.1103/PhysRevE.54.5286} {\bibfield  {journal} {\bibinfo
  {journal} {Physical Review E}\ }\textbf {\bibinfo {volume} {54}},\ \bibinfo
  {pages} {5286} (\bibinfo {year} {1996})}\BibitemShut {NoStop}%
\bibitem [{\citenamefont {Sch{\"{a}}tzel}\ and\ \citenamefont
  {Ackerson}(1993)}]{schaetzel1993}%
  \BibitemOpen
  \bibfield  {author} {\bibinfo {author} {\bibfnamefont {K.}~\bibnamefont
  {Sch{\"{a}}tzel}}\ and\ \bibinfo {author} {\bibfnamefont {B.~J.}\
  \bibnamefont {Ackerson}},\ }\href {\doibase 10.1103/PhysRevE.48.3766}
  {\bibfield  {journal} {\bibinfo  {journal} {Physical Review E}\ }\textbf
  {\bibinfo {volume} {48}},\ \bibinfo {pages} {3766} (\bibinfo {year}
  {1993})}\BibitemShut {NoStop}%
\bibitem [{\citenamefont {Sinn}\ \emph {et~al.}(2001)\citenamefont {Sinn},
  \citenamefont {Heymann}, \citenamefont {Stipp},\ and\ \citenamefont
  {Palberg}}]{Sinn2001}%
  \BibitemOpen
  \bibfield  {author} {\bibinfo {author} {\bibfnamefont {C.}~\bibnamefont
  {Sinn}}, \bibinfo {author} {\bibfnamefont {A.}~\bibnamefont {Heymann}},
  \bibinfo {author} {\bibfnamefont {A.}~\bibnamefont {Stipp}}, \ and\ \bibinfo
  {author} {\bibfnamefont {T.}~\bibnamefont {Palberg}},\ }\href {\doibase
  https://doi.org/10.1007/3-540-45725-9_57} {\bibfield  {journal} {\bibinfo
  {journal} {Trends in Colloid and Interface Science XV. Progress in Colloid
  and Polymer Science}\ }\textbf {\bibinfo {volume} {118}},\ \bibinfo {pages}
  {266} (\bibinfo {year} {2001})}\BibitemShut {NoStop}%
\bibitem [{\citenamefont {Filion}\ \emph
  {et~al.}(2010{\natexlab{b}})\citenamefont {Filion}, \citenamefont {Hermes},
  \citenamefont {Ni},\ and\ \citenamefont {Dijkstra}}]{Filion2010a}%
  \BibitemOpen
  \bibfield  {author} {\bibinfo {author} {\bibfnamefont {L.}~\bibnamefont
  {Filion}}, \bibinfo {author} {\bibfnamefont {M.}~\bibnamefont {Hermes}},
  \bibinfo {author} {\bibfnamefont {R.}~\bibnamefont {Ni}}, \ and\ \bibinfo
  {author} {\bibfnamefont {M.}~\bibnamefont {Dijkstra}},\ }\href {\doibase
  10.1063/1.3506838} {\bibfield  {journal} {\bibinfo  {journal} {Journal of
  Chemical Physics}\ }\textbf {\bibinfo {volume} {133}},\ \bibinfo {pages}
  {244115} (\bibinfo {year} {2010}{\natexlab{b}})}\BibitemShut {NoStop}%
\bibitem [{\citenamefont {Fiorucci}\ \emph
  {et~al.}(2020{\natexlab{b}})\citenamefont {Fiorucci}, \citenamefont {Coli},
  \citenamefont {Padding},\ and\ \citenamefont {Dijkstra}}]{Fiorucci2020a}%
  \BibitemOpen
  \bibfield  {author} {\bibinfo {author} {\bibfnamefont {G.}~\bibnamefont
  {Fiorucci}}, \bibinfo {author} {\bibfnamefont {G.~M.}\ \bibnamefont {Coli}},
  \bibinfo {author} {\bibfnamefont {J.~T.}\ \bibnamefont {Padding}}, \ and\
  \bibinfo {author} {\bibfnamefont {M.}~\bibnamefont {Dijkstra}},\ }\href
  {\doibase 10.1063/1.5137815} {\bibfield  {journal} {\bibinfo  {journal}
  {Journal of Chemical Physics}\ }\textbf {\bibinfo {volume} {152}},\ \bibinfo
  {pages} {064903} (\bibinfo {year} {2020}{\natexlab{b}})}\BibitemShut
  {NoStop}%
\bibitem [{\citenamefont {Bannerman}\ \emph {et~al.}(2014)\citenamefont
  {Bannerman}, \citenamefont {Strobl}, \citenamefont {Formella},\ and\
  \citenamefont {P{\"{o}}schel}}]{Bannerman2014}%
  \BibitemOpen
  \bibfield  {author} {\bibinfo {author} {\bibfnamefont {M.~N.}\ \bibnamefont
  {Bannerman}}, \bibinfo {author} {\bibfnamefont {S.}~\bibnamefont {Strobl}},
  \bibinfo {author} {\bibfnamefont {A.}~\bibnamefont {Formella}}, \ and\
  \bibinfo {author} {\bibfnamefont {T.}~\bibnamefont {P{\"{o}}schel}},\ }\href
  {\doibase 10.1007/s40571-014-0021-8} {\bibfield  {journal} {\bibinfo
  {journal} {Computational Particle Mechanics}\ }\textbf {\bibinfo {volume}
  {1}},\ \bibinfo {pages} {191} (\bibinfo {year} {2014})}\BibitemShut {NoStop}%
\bibitem [{\citenamefont {Steinhardt}\ \emph {et~al.}(1983)\citenamefont
  {Steinhardt}, \citenamefont {Nelson},\ and\ \citenamefont
  {Ronchetti}}]{Steinhardt1983}%
  \BibitemOpen
  \bibfield  {author} {\bibinfo {author} {\bibfnamefont {P.~J.}\ \bibnamefont
  {Steinhardt}}, \bibinfo {author} {\bibfnamefont {D.~R.}\ \bibnamefont
  {Nelson}}, \ and\ \bibinfo {author} {\bibfnamefont {M.}~\bibnamefont
  {Ronchetti}},\ }\href {\doibase 10.1103/PhysRevB.28.784} {\bibfield
  {journal} {\bibinfo  {journal} {Physical Review B}\ }\textbf {\bibinfo
  {volume} {28}},\ \bibinfo {pages} {784} (\bibinfo {year} {1983})}\BibitemShut
  {NoStop}%
\bibitem [{\citenamefont {{Ten Wolde}}\ \emph {et~al.}(1995)\citenamefont {{Ten
  Wolde}}, \citenamefont {Ruiz-Montero},\ and\ \citenamefont
  {Frenkel}}]{TenWolde1995}%
  \BibitemOpen
  \bibfield  {author} {\bibinfo {author} {\bibfnamefont {P.~R.}\ \bibnamefont
  {{Ten Wolde}}}, \bibinfo {author} {\bibfnamefont {M.~J.}\ \bibnamefont
  {Ruiz-Montero}}, \ and\ \bibinfo {author} {\bibfnamefont {D.}~\bibnamefont
  {Frenkel}},\ }\href {\doibase 10.1103/PhysRevLett.75.2714} {\bibfield
  {journal} {\bibinfo  {journal} {Physical Review Letters}\ }\textbf {\bibinfo
  {volume} {75}},\ \bibinfo {pages} {2714} (\bibinfo {year}
  {1995})}\BibitemShut {NoStop}%
\bibitem [{\citenamefont {Deemer}\ and\ \citenamefont
  {Votaw}(1955)}]{Deemer1955}%
  \BibitemOpen
  \bibfield  {author} {\bibinfo {author} {\bibfnamefont {W.~L.}\ \bibnamefont
  {Deemer}}\ and\ \bibinfo {author} {\bibfnamefont {D.~F.}\ \bibnamefont
  {Votaw}},\ }\href {\doibase 10.1214/aoms/1177728494} {\bibfield  {journal}
  {\bibinfo  {journal} {The Annals of Mathematical Statistics}\ }\textbf
  {\bibinfo {volume} {26}},\ \bibinfo {pages} {498} (\bibinfo {year}
  {1955})}\BibitemShut {NoStop}%
\bibitem [{\citenamefont {Bolhuis}\ and\ \citenamefont
  {Kofke}(1996)}]{Bolhuis1996}%
  \BibitemOpen
  \bibfield  {author} {\bibinfo {author} {\bibfnamefont {P.~G.}\ \bibnamefont
  {Bolhuis}}\ and\ \bibinfo {author} {\bibfnamefont {D.~A.}\ \bibnamefont
  {Kofke}},\ }\href {\doibase 10.1103/PhysRevE.54.634} {\bibfield  {journal}
  {\bibinfo  {journal} {Physical Review E - Statistical Physics, Plasmas,
  Fluids, and Related Interdisciplinary Topics}\ }\textbf {\bibinfo {volume}
  {54}},\ \bibinfo {pages} {634} (\bibinfo {year} {1996})}\BibitemShut
  {NoStop}%
\bibitem [{\citenamefont {Stukowski}(2010)}]{ovito}%
  \BibitemOpen
  \bibfield  {author} {\bibinfo {author} {\bibfnamefont {A.}~\bibnamefont
  {Stukowski}},\ }\href {\doibase 10.1088/0965-0393/18/1/015012} {\bibfield
  {journal} {\bibinfo  {journal} {{MODELLING AND SIMULATION IN MATERIALS
  SCIENCE AND ENGINEERING}}\ }\textbf {\bibinfo {volume} {{18}}} (\bibinfo
  {year} {{2010}}),\ 10.1088/0965-0393/18/1/015012}\BibitemShut {NoStop}%
\bibitem [{\citenamefont {O’Malley}\ and\ \citenamefont
  {Snook}(2003)}]{omalley2003}%
  \BibitemOpen
  \bibfield  {author} {\bibinfo {author} {\bibfnamefont {B.}~\bibnamefont
  {O’Malley}}\ and\ \bibinfo {author} {\bibfnamefont {I.}~\bibnamefont
  {Snook}},\ }\href@noop {} {\bibfield  {journal} {\bibinfo  {journal}
  {Physical Review Letters}\ }\textbf {\bibinfo {volume} {90}},\ \bibinfo
  {pages} {085702} (\bibinfo {year} {2003})}\BibitemShut {NoStop}%
\bibitem [{\citenamefont {Harland}\ \emph {et~al.}(1995)\citenamefont
  {Harland}, \citenamefont {Henderson}, \citenamefont {Underwood},\ and\
  \citenamefont {van Megen}}]{harland1995}%
  \BibitemOpen
  \bibfield  {author} {\bibinfo {author} {\bibfnamefont {J.}~\bibnamefont
  {Harland}}, \bibinfo {author} {\bibfnamefont {S.}~\bibnamefont {Henderson}},
  \bibinfo {author} {\bibfnamefont {S.~M.}\ \bibnamefont {Underwood}}, \ and\
  \bibinfo {author} {\bibfnamefont {W.}~\bibnamefont {van Megen}},\ }\href@noop
  {} {\bibfield  {journal} {\bibinfo  {journal} {Physical review letters}\
  }\textbf {\bibinfo {volume} {75}},\ \bibinfo {pages} {3572} (\bibinfo {year}
  {1995})}\BibitemShut {NoStop}%
\bibitem [{\citenamefont {Sch{\"o}pe}\ \emph {et~al.}(2006)\citenamefont
  {Sch{\"o}pe}, \citenamefont {Bryant},\ and\ \citenamefont
  {Van~Megen}}]{schope2006}%
  \BibitemOpen
  \bibfield  {author} {\bibinfo {author} {\bibfnamefont {H.~J.}\ \bibnamefont
  {Sch{\"o}pe}}, \bibinfo {author} {\bibfnamefont {G.}~\bibnamefont {Bryant}},
  \ and\ \bibinfo {author} {\bibfnamefont {W.}~\bibnamefont {Van~Megen}},\
  }\href@noop {} {\bibfield  {journal} {\bibinfo  {journal} {Physical Review
  Letters}\ }\textbf {\bibinfo {volume} {96}},\ \bibinfo {pages} {175701}
  (\bibinfo {year} {2006})}\BibitemShut {NoStop}%
\bibitem [{\citenamefont {Iacopini}\ \emph {et~al.}(2009)\citenamefont
  {Iacopini}, \citenamefont {Palberg},\ and\ \citenamefont
  {Sch{\"o}pe}}]{iacopini2009}%
  \BibitemOpen
  \bibfield  {author} {\bibinfo {author} {\bibfnamefont {S.}~\bibnamefont
  {Iacopini}}, \bibinfo {author} {\bibfnamefont {T.}~\bibnamefont {Palberg}}, \
  and\ \bibinfo {author} {\bibfnamefont {H.~J.}\ \bibnamefont {Sch{\"o}pe}},\
  }\href@noop {} {\bibfield  {journal} {\bibinfo  {journal} {The Journal of
  chemical physics}\ }\textbf {\bibinfo {volume} {130}},\ \bibinfo {pages}
  {084502} (\bibinfo {year} {2009})}\BibitemShut {NoStop}%
\bibitem [{\citenamefont {Sch{\"a}tzel}\ and\ \citenamefont
  {Ackerson}(1993)}]{schatzel1993a}%
  \BibitemOpen
  \bibfield  {author} {\bibinfo {author} {\bibfnamefont {K.}~\bibnamefont
  {Sch{\"a}tzel}}\ and\ \bibinfo {author} {\bibfnamefont {B.~J.}\ \bibnamefont
  {Ackerson}},\ }\href@noop {} {\bibfield  {journal} {\bibinfo  {journal}
  {Physica Scripta}\ }\textbf {\bibinfo {volume} {1993}},\ \bibinfo {pages}
  {70} (\bibinfo {year} {1993})}\BibitemShut {NoStop}%
\bibitem [{\citenamefont {Gasser}\ \emph {et~al.}(2001)\citenamefont {Gasser},
  \citenamefont {Weeks}, \citenamefont {Schofield}, \citenamefont {Pusey},\
  and\ \citenamefont {Weitz}}]{gasser2001}%
  \BibitemOpen
  \bibfield  {author} {\bibinfo {author} {\bibfnamefont {U.}~\bibnamefont
  {Gasser}}, \bibinfo {author} {\bibfnamefont {E.~R.}\ \bibnamefont {Weeks}},
  \bibinfo {author} {\bibfnamefont {A.}~\bibnamefont {Schofield}}, \bibinfo
  {author} {\bibfnamefont {P.}~\bibnamefont {Pusey}}, \ and\ \bibinfo {author}
  {\bibfnamefont {D.}~\bibnamefont {Weitz}},\ }\href@noop {} {\bibfield
  {journal} {\bibinfo  {journal} {Science}\ }\textbf {\bibinfo {volume}
  {292}},\ \bibinfo {pages} {258} (\bibinfo {year} {2001})}\BibitemShut
  {NoStop}%
\bibitem [{\citenamefont {Kolmogorow}(1937)}]{Kolmogorow1937}%
  \BibitemOpen
  \bibfield  {author} {\bibinfo {author} {\bibfnamefont {A.}~\bibnamefont
  {Kolmogorow}},\ }\href {http://mi.mathnet.ru/izv3359} {\bibfield  {journal}
  {\bibinfo  {journal} {Izv. Akad. Nauk SSSR Ser. Mat.}\ }\textbf {\bibinfo
  {volume} {1}},\ \bibinfo {pages} {355 } (\bibinfo {year} {1937})}\BibitemShut
  {NoStop}%
\bibitem [{\citenamefont {Avrami}(1939)}]{Avrami1939}%
  \BibitemOpen
  \bibfield  {author} {\bibinfo {author} {\bibfnamefont {M.}~\bibnamefont
  {Avrami}},\ }\href {\doibase 10.1063/1.1750380} {\bibfield  {journal}
  {\bibinfo  {journal} {The Journal of Chemical Physics}\ }\textbf {\bibinfo
  {volume} {7}},\ \bibinfo {pages} {1103} (\bibinfo {year} {1939})}\BibitemShut
  {NoStop}%
\bibitem [{\citenamefont {Avrami}(1940)}]{Avrami1940}%
  \BibitemOpen
  \bibfield  {author} {\bibinfo {author} {\bibfnamefont {M.}~\bibnamefont
  {Avrami}},\ }\href {\doibase 10.1063/1.1750631} {\bibfield  {journal}
  {\bibinfo  {journal} {The Journal of Chemical Physics}\ }\textbf {\bibinfo
  {volume} {8}},\ \bibinfo {pages} {212} (\bibinfo {year} {1940})}\BibitemShut
  {NoStop}%
\bibitem [{\citenamefont {Avrami}(1941)}]{Avrami1941}%
  \BibitemOpen
  \bibfield  {author} {\bibinfo {author} {\bibfnamefont {M.}~\bibnamefont
  {Avrami}},\ }\href {\doibase 10.1063/1.1750872} {\bibfield  {journal}
  {\bibinfo  {journal} {The Journal of Chemical Physics}\ }\textbf {\bibinfo
  {volume} {9}},\ \bibinfo {pages} {177} (\bibinfo {year} {1941})}\BibitemShut
{NoStop}%
	\bibitem [{\citenamefont {Stukowski}(2010)}]{ovito2010}%
  \BibitemOpen
  \bibfield  {author} {\bibinfo {author} {\bibfnamefont {A.}~\bibnamefont
  {Stukowski}},\ }\href {\doibase 10.1088/0965-0393/18/1/015012} {\bibfield  {journal}
  {\bibinfo  {journal} {Modelling and simulation in materials science and engineering}\ }\textbf {\bibinfo
  {volume} {18}},\ \bibinfo {number} {1} (\bibinfo {year} {2010})}\BibitemShut
  {NoStop}%
\end{thebibliography}
\end{document}